\documentclass{lmcs}
\pdfoutput=1

\usepackage{lastpage}
\lmcsdoi{20}{1}{21}
\lmcsheading{}{\pageref{LastPage}}{}{}%
{Feb.~05,~2022}{Mar.~08,~2024}{}

\pdfoutput=1 % arXiv

\usepackage[utf8]{inputenc}

\usepackage{amsmath,amssymb}
\usepackage{amsthm}
\usepackage{mathtools} %\coloneqq
\usepackage{stmaryrd} %\bigsqcup

\usepackage{microtype}

\usepackage{subcaption}
\usepackage{tikz}
\usetikzlibrary{arrows,calc,backgrounds,shapes,patterns,positioning,fit}
\usepackage{booktabs}

\usepackage{hyperref} % must come before clerever
\usepackage[capitalize]{cleveref}

\newcommand{\renewtheorem}[1]{%
  \expandafter\let\csname #1\endcsname\relax
  \expandafter\let\csname c@#1\endcsname\relax
  \expandafter\let\csname end#1\endcsname\relax
  \newtheorem{#1}%
}

\theoremstyle{plain}
\renewtheorem{thm}{Theorem}[section]
\renewtheorem{cor}[thm]{Corollary}
\renewtheorem{lem}[thm]{Lemma}
\renewtheorem{prop}[thm]{Proposition}

\theoremstyle{definition}
\renewtheorem{exa}[thm]{Example}
\renewtheorem{defi}[thm]{Definition}

\newcommand{\cornerhere}{\hfill$\lrcorner$\gdef\ExampleEndMarker{}}
\newenvironment{example}{\gdef\ExampleEndMarker{\hfill$\lrcorner$}\exa}{\ExampleEndMarker\endexa}

\newenvironment{aquote}[1]{\def\quoteauthor{#1}\quote\itshape}{\hfill\textsc{--- \quoteauthor}\endquote\vspace{.5\normalbaselineskip}}

\newcommand{\E}{\exists}
\newcommand{\A}{\forall}
\renewcommand{\phi}{\varphi}
\renewcommand{\theta}{\vartheta}
\renewcommand{\emptyset}{\varnothing}
\newcommand*{\ext}[1]{[\![ #1 ]\!]}
\newcommand{\LFP}{{\rm LFP}}
\DeclareMathOperator{\lfp}{\mathbf{lfp}}
\DeclareMathOperator{\gfp}{\mathbf{gfp}}
\DeclareMathOperator{\nnf}{\mathrm{nnf}}
\DeclareMathOperator{\Lit}{\mathrm{Lit}}
\newcommand{\lit}{\alpha} % variable for arbitrary literals

\newcommand{\from}{\colon}
\newcommand{\dcup}{\dot\cup}
\newcommand*{\tup}[1]{\mathbf{#1}}
\newcommand{\ta}{\tup a}
\newcommand{\tb}{\tup b}
\newcommand{\tx}{\tup x}
\newcommand{\ty}{\tup y}
\newcommand{\co}{\colon}

\renewcommand{\AA}{{\mathfrak A}}
\newcommand{\N}{{\mathbb N}} % natural numbers
\newcommand{\K}{{\mathbb K}} % semiring
\newcommand{\Gg}{\mathcal{G}} % game
\newcommand{\Ss}{\mathcal{S}} % strategy
\newcommand{\Mm}{{\mathcal M}} % strategy in MC
\newcommand{\Tt}{{\mathsf T}} % tree unraveling

\DeclareMathOperator{\Strat}{\mathrm{Strat}}
\DeclareMathOperator{\WinStrat}{\mathrm{WinStrat}}

\newcommand{\Bool}{\mathbb{B}}
\newcommand{\Nat}{\mathbb{N}}
\newcommand{\Sinf}{{\mathbb S}^{\infty}}
\newcommand{\Trop}{\mathbb{T}}
\newcommand{\Vit}{\mathbb{V}}
\newcommand{\PosBool}{\mathsf{PosBool}}

\renewcommand{\bar}{\overline}
\newcommand*{\ps}[1]{[\![ #1 ]\!]}
\newcommand*{\nn}[1]{\bar{#1}}
\newcommand{\nnX}{\nn{X}}
\newcommand{\nnx}{\nn{x}}

\newcommand{\Inf}{\bigsqcap}
\newcommand{\Sup}{\bigsqcup}
\newcommand{\Bone}{\mathbf{1}}
\newcommand{\Bzero}{\mathbf{0}}
\newcommand{\bcdot}{\boldsymbol\cdot} % bigger \cdot (easier to see in text)
\newcommand{\FormulaWin}{\mathsf{win}_0}

\newcommand{\absorb}{\succeq}
\newcommand{\absorbneq}{\succ}

\newcommand{\var}{\mathsf{var}}

\newcommand{\pitrack}{\pi_{\text{\sffamily\upshape strat}}}
\newcommand{\pirev}{\pi_{\text{\sffamily\upshape repair}}}
\newcommand{\piwin}{\pi_{\text{\sffamily\upshape target}}}

\newcommand{\prefix}{\sqsubseteq}
\newcommand{\ecount}[2]{\#_{#2}(#1)} %number of times the edge #2 occurs in strategy #1
\newcommand{\ep}[1]{\#_E(#1)} % edge profile

\tikzset{
    arr/.style={draw,->,>=stealth',shorten <=2pt,shorten >=2pt,every node/.style={auto,inner sep=2pt,font=\scriptsize}},
    gamenode/.style={draw,inner sep=2pt,minimum size=.4cm},
    p0/.style={gamenode,circle},
    p1/.style={gamenode,rectangle},
    F/.style={thick, pattern=north east lines},
    dot/.style={circle,draw,fill,black,minimum size=3pt,inner sep=0pt},
    marker/.style={draw=none,inner sep=0pt,overlay},
    short/.style={ shorten >=#1, shorten <=#1 }
}

\begin{document}

\title[Strategy Analysis in Büchi Games by Semiring Valuations]{Semiring Provenance for Büchi Games:\texorpdfstring{\\}{ }Strategy Analysis with Absorptive Polynomials}
\titlecomment{This is a revised and expanded version of the conference paper \cite{GraedelLucNaa21}, presented at GandALF 2021.}

\author[E.~Grädel]{Erich Grädel\lmcsorcid{0000-0002-8950-9991}}
\author[N.~Lücking]{Niels Lücking}
\author[M.~Naaf]{Matthias Naaf\lmcsorcid{0000-0002-1099-5713}}

\address{RWTH Aachen University, Germany}
\email{graedel@logic.rwth-aachen.de, niels.luecking@rwth-aachen.de, naaf@logic.rwth-aachen.de}

\keywords{Büchi games, strategy analysis, semiring semantics, provenance analysis, fixed-point logic}

%%%%%%%%%%%%%%%%%%%%%%%%%%%%%%%%
%%  ABSTRACT
%%%%%%%%%%%%%%%%%%%%%%%%%%%%%%%%

\begin{abstract}
This paper presents a case study for the application of semiring semantics for fixed-point formulae
to the analysis of strategies in Büchi games.
Semiring semantics generalizes the classical Boolean semantics by permitting multiple truth values from certain semirings.
Evaluating the fixed-point formula that defines the winning region in a given game in an appropriate
semiring of polynomials provides not only the Boolean information on \emph{who} wins, but also tells us \emph{how} they win and \emph{which strategies} they might use. This is well-understood for reachability games, where the winning region is definable as a least fixed point.
The case of Büchi games is of special interest, not only due to their practical importance, but also because it is the simplest case where
the fixed-point definition involves a genuine alternation of a greatest and a least fixed point.

We show that, in a precise sense, semiring semantics provide information about all \emph{absorption-dominant} strategies -- strategies that win with minimal effort,
and we discuss how these relate to positional and the more general \emph{persistent} strategies.
This information enables applications such as game synthesis or determining minimal modifications to the game needed to change its outcome.
Lastly, we discuss limitations of our approach and present questions that cannot be immediately answered by semiring semantics.
\end{abstract}

\maketitle

%%%%%%%%%%%%%%%%%%%%%%%%%%%%%%%%
%%  MAIN BODY
%%%%%%%%%%%%%%%%%%%%%%%%%%%%%%%%

\section{Introduction}

Two-player games on finite graphs which admit infinite plays are of fundamental importance in many areas of logic and computer science, especially in the 
formal analysis of reactive systems, where they model the non-terminating interaction between a system and its environment.
In such a game, the \emph{objective} or \emph{winning condition} of the player who represents the system specifies
the desired set of behaviours of the system. The most basic classes of such
objectives are \emph{reachability} and \emph{safety} objectives defined by a set of states (positions) that the player should reach, or avoid.
We can assume, without loss of generality, that even though infinite plays are possible in a game with reachability or safety objectives, 
they are all won by the same player.

Games with genuine and non-trivial winning conditions for infinite plays are harder to analyse; they include games with arbitrary
$\omega$-regular objectives, such as liveness, Muller, Streett-Rabin, or parity objectives, and many others. 
The goal of this paper is to provide a case study of a recent method for strategy analysis, based on semiring semantics,
and we would like to explore its potential for providing detailed information about strategies in genuinely infinite games.  
One of the simplest class of games with a non-trivial winning condition for the infinite plays are games with 
Büchi objectives, which require that a specific target set $F$  of states is reached infinitely often during the play (see e.g. \cite{GraedelThoWil02} for background). 
Büchi games, as well as some of their straightforward generalisations, have many applications in formal methods,
and efficient algorithms for solving them have been studied thoroughly (see e.g. \cite{ChatterjeeDvoHenLoi16,ChatterjeeHen14,ChatterjeeHenPit08}).
They are also of interest from the points of view of topology and logic, because they are among the simplest games where the set of winning plays is neither open nor closed, and where logical definition of the winning region requires a genuine alternation of a greatest and a least fixed point (see \cref{sec:TheFormula}).
 
Strategies in infinite games can be very complicated because, in principle, they may depend on the entire history of a play.
Thus, there exist uncountably many different strategies, even on a finite game graph.
Fortunately, in many cases and in particular for Büchi games, simple strategies are sufficient to win.
A fundamental result in this context is the positional determinacy of parity games (of which Büchi games are a special case),
saying that from each position, one of the two players has a \emph{positional winning strategy}, i.e.
a winning strategy that only depends on the current position and not on the history of the play.   
A positional strategy can be viewed as a subgraph of the game graph, and can therefore be represented in
a compact way. As a consequence, the algorithmic analysis of Büchi games has concentrated almost exclusively on the positional strategies. Here we extend this point of view somewhat and take also other kinds of simple strategies into account.  
Specifically, we are interested in \emph{absorption-dominant} winning strategies \cite{GraedelTan20} which are strategies without redundant
moves; this means that taking away anything, in the sense of demanding that some specific move is played less often,
makes the strategy non-winning.
Another way to distinguish positional strategies from absorption-dominant ones concerns their minimisation properties:
while positional strategies minimize the \emph{set} of moves that they use, absorption-dominant strategies
take multiplicities into account and minimize the \emph{multiset} of moves.  
A further interesting class are the \emph{persistent} strategies \cite{MarcinkowskiTruderung02}, which are positional in each individual play
but not necessarily across distinct plays.
We shall study the relationship between these different classes of simple strategies, and prove that
every positional strategy is absorption-dominant and every absorption-dominant strategy is persistent,
and that these inclusions are strict.

The specific method for strategy analysis that we want to apply to Büchi games in this paper is
based on the logical definability of the winning positions by 
a formula in the fixed-point logic LFP, and on the semiring semantics for LFP
developed in \cite{DannertGraNaaTan21}.
In the classical Boolean semantics, a model $\AA$ of a formula $\phi$
assigns to each (instantiated) literal a Boolean value.
$\K$-interpretations $\pi$, for a suitable semiring $\K$, generalize this
by assigning to each such literal a semiring value from $\K$.
We then interpret $0$ as \emph{false} and all other semiring values as \emph{nuances of true}
that provide additional information, depending on the semiring:
For example, the Boolean semiring $\Bool = (\{0,1\}, \lor, \land, 0, 1)$ corresponds to Boolean semantics, the Viterbi-semiring $\Vit = ([0,1], \max, \cdot, 0, 1)$ can model \emph{confidence} scores, 
the tropical semiring $\Trop= (\mathbb{R}_{+}^{\infty},\min,+,\infty,0)$
is used for cost analysis, and min-max-semirings $(A, \max, \min, a, b)$ for a totally ordered set $(A,<)$ can model different access levels.
Most importantly, semirings of polynomials, such as $\N[X]$, allow us to \emph{track} certain literals by mapping them to different indeterminates. The overall value of the formula is then a polynomial that describes precisely what combinations of literals prove the truth of the formula.
Semiring semantics has been studied for various logics \cite{BourgauxOzaPenPre20, DannertGra19, DannertGra20,DannertGraNaaTan21,GraedelTan17},
following the successful development of semiring provenance in database theory and related fields (see e.g.\cite{DeutchMilRoyTan14,GeertsPog10,GreenKarTan07,GreenTan17,OzakiPen18,RaghothamanMenZhaNaiSch20,Senellart17}).
While semiring provenance analysis for database queries had largely been  confined
to positive query languages such as conjunctive queries, positive relational algebra, and Datalog,
the generalisation to logics such as first-order logic FO and least fixed-point logic LFP -- featuring full negation and unrestricted interaction between least and greatest fixed points -- poses non-trivial mathematical  challenges and requires  
new algebraic constructions. Specifically, it has turned out that appropriate semirings for LFP should be
absorptive and fully continuous \cite{DannertGraNaaTan21}.
Fortunately, this is the case for most of the important application semirings such
as $\Vit, \Trop$ or min-max-semirings, but not for the natural semiring $\N$, or the general provenance semirings
of polynomials or formal power series, $\N[X]$ and $\N^\infty \ps X$. 
Instead, we rely here on an absorptive version of $\N[X]$ (and $\N^\infty \ps X$),
the semirings $\Sinf[X]$ of \emph{generalized absorptive polynomials}, which we 
discuss in \cref{sect:semirings}.
These are the \emph{universal} absorptive, fully-continuous semirings, in the sense that 
every mapping $h \colon X \to \K$ into an absorptive, fully-continuous semiring $\K$
uniquely extends to a fully-continuous semiring homomorphism $h \colon \Sinf[X] \to \K$ that preserves LFP semantics.
From valuations of fixed-point formulae in such semirings we thus can 
derive detailed insights into why the formula holds -- and by applying this to
the fixed-point definition of winning positions in Büchi games we obtain compact descriptions of 
winning strategies, in particular of all positional strategies and all absorption-dominant ones.

After an analysis of simple winning strategies in Büchi games, and a short introduction to semiring semantics for fixed-point logic, we shall study the semiring valuations of the particular LFP-formula $\FormulaWin(x)$
that defines the winning region for Player~0 in Büchi games. Given that the objective of Player~0 is to ensure 
that the play hits the target set $F$
infinitely often, we may informally describe their winning region  as the \emph{largest} set
$Y$ of positions from which they can enforce a (further) visit to $Y\cap F$
after $k\geq 1$ moves.  On the other side the set of positions from which
Player~0 can enforce a visit to a target set is the \emph{smallest} set 
of positions that either are already in the target set, or from which Player~0 can 
enforce the play to come closer to it. Thus, the winning region of Player~0 
can be described as a greatest fixed point inside of which there is a least fixed point, and it is well-known that 
this fixed-point alternation in the treatment of Büchi objectives cannot be avoided, see e.g. \cite{BradfieldWal18}.

We shall prove a Sum-of-Strategies Theorem, saying that for any position $v$ in a Büchi game, the
valuation of the LFP-formula $\FormulaWin(v)$ in an absorptive, fully-continuous semiring
coincides with the sum of the valuations of all absorption-dominant winning strategies from $v$.
Besides being of theoretical interest, this result allows to study a number of interesting questions
concerning the available winning strategies in a given Büchi game:

\smallskip\noindent\textbf{Strategy tracking.}
Introducing indeterminates for all edges in a fixed Büchi game $\Gg$,
the semiring value $\pitrack \ext {\FormulaWin(v)}$ for a position $v$
is a polynomial whose monomials are concise descriptions of all absorption-dominant strategies.
From these monomials we can derive whether Player 0 wins from $v$ (if there are any monomials)
and which edges are used by each absorption-dominant strategy, and how often they appear in the strategy tree.
In particular, we can immediately identify and count positional strategies from the polynomial.
Going further, we can answer questions such as: can Player~0 still win if we remove edge $e$, or several edges at once?
Can they still win if edge $e$ may only be used finitely often in each play?

\smallskip\noindent\textbf{Repairing a game.} Instead of analysing strategies in a fixed game, we may also reason about modifications or synthesis of (parts of) the game.
For example, assuming Player~0 loses from $v$, what are minimal modifications to the game that would let Player~0 win from $v$?
To answer such questions we have to take into account also negative information (i.e., absent edges in the graph),
so as to find a minimal repair consisting of both moves to delete and moves to add.
Algebraically, this requires to extend our semirings by dual-indeterminates, which leads to quotient semirings
$\Sinf[X,\bar X]$ by a construction that has been used before in \cite{GraedelTan17,XuZhaAlaTan18,DannertGraNaaTan21} to deal with
semiring semantics for negation.
We illustrate with the example of minimal repairs that we can indeed derive the desired information from valuations in such semirings.

\smallskip\noindent\textbf{Cost computation.}  A typical application of semiring provenance in databases is cost analysis:
assuming that atomic facts are not for free but come with a cost (a non-negative real number), 
then the minimal cost of evaluating a query is described by a provenance valuation in
the tropical semiring $\Trop= (\mathbb{R}_{+}^{\infty},\min,+,\infty,0)$. In a game, we may ask
the analogous question of what is the minimal cost of a winning strategy assuming that  
moves come with a cost. For reachability and safety games that admit only finite plays, such an analysis
works in a reasonably straightforward way by means of an appropriate 
sum-of-strategies theorem \cite{GraedelTan20} (which is much simpler than the one for Büchi games).
However, as we shall show, this does not generalize in a nice way 
to Büchi games, and this seems to be a general limitation of the method of semiring valuations.

\section{Büchi Games and Strategies}

A Büchi game is given by a tuple $\Gg = (V, V_0, V_1, E, F)$ where
$V$ is a set of positions (here assumed to be finite), with a disjoint decomposition
$V=V_0 \dcup V_1$ into positions of Player~0 and positions of Player~1.
The relation $E\subseteq V\times V$ specifies the possible moves, and the target set
$F\subseteq V$ describes the winning condition.
We denote the set of immediate successors of a position $v$ by 
$vE:=\{w \mid vw\in E\}$ and require that $vE\neq\emptyset$ for all $v$.
A play from an initial position $v_0$ is an infinite path $v_0v_1v_2\dots$
through $\Gg$ where the successor $v_{i+1}\in v_iE$ is chosen by Player~0
if $v_i\in V_0$ and by Player~1 if $v_1\in V_1$.
A play $v_0v_1v_2\dots$ is won by Player~0 if $v_i\in F$ for infinitely many $i<\omega$,
otherwise it is won by Player~1. 
The winning region of Player~$\sigma$ is the set
of those positions $v\in V$ such that Player~$\sigma$ has a winning strategy
from $v$, i.e. a strategy that guarantees them a win, no matter what the opponent does.

A strategy for Player~$\sigma$ in $\Gg = (V, V_0, V_1, E, F)$ can be represented in different ways,
for instance as a function $f \co V^* V_\sigma \to V$ that assigns a next position to each partial play ending in a position of Player $\sigma$,
or simply $f \co V_\sigma \to V$ if the strategy is positional.
Here we follow an alternative approach and represent strategies as trees,
comprised of all plays that are consistent with the strategy (see, e.g., \cite{GraedelTan20}).
For simplicity, we only consider strategies of Player~0, so unless mentioned otherwise, \emph{strategy} always refers to a strategy for Player~0.

\begin{defi}
Given a Büchi game $\Gg = (V, V_0, V_1, E, F)$, the \emph{tree unraveling} from $v_0$ is the tree $\Tt(\Gg, v_0)$ whose nodes are all finite paths $\rho$ from $v_0$ in $\Gg$ and whose edges are $\rho \to \rho v$ for $v \in V$.
We often denote a node of $\Tt(\Gg, v_0)$ as $\rho v$ to indicate a finite path ending in $v \in V$.
We write $|\rho|$ for the length of $\rho$ and $\rho \prefix \rho'$ if $\rho$ is a (not necessarily strict) prefix of $\rho'$.
\end{defi}

Strategies can then be defined as subtrees of the tree unraveling, which allows for a more visual way to reason about strategies.
An important detail is that the strategy tree only contains positions (and thus choices for these positions) that 
are reachable when following the strategy.
Moreover, we only consider finite Büchi games and hence the tree unraveling and all strategies are finitely branching. 

\begin{defi}
\label{def:strategyAsTree}
A \emph{strategy} $\Ss$ (of Player~0) from $v_0$ in $\Gg$ is a subtree of $\Tt(\Gg, v_0)$ induced by a node set $W$ satisfying the following conditions:
\begin{itemize}
\item if $\rho v \in W$, then also $\rho \in W$ (prefix closure);
\item if $\rho v \in W$ and $v \in V_0$, then there is a unique $v' \in vE$ with $\rho v v' \in W$ (unique choice);
\item if $\rho v \in W$ and $v \in V_1$, then $\rho v v' \in W$ for all $v' \in vE$ (all moves of the opponent).
\end{itemize}
The strategy is winning if all plays contained in $\Ss$ are winning.

We commonly write $\rho \in \Ss$ instead of $\rho \in W$,
and we often refer to paths of the form $\rho v \in \Ss$ as \emph{occurrences of $v$} in $\Ss$.
When we depict strategies graphically, we represent finite paths $\rho v$ just by their last position $v$ to ease readability (notice that in the tree unravelling 
$\rho$ can be reconstructed from $v$ by following the path to the root).
See \cref{fig:RunningStrategy} for an example.
For $v \in V_0$, we further write $\Ss(\rho v) = w$ if $\rho v w$ is the (unique) successor of $\rho v$ in $\Ss$.
If $\Ss$ is positional, we may also write $\Ss(v)$ to denote the unique successor of $v$ chosen by $\Ss$.
We write $\Strat_\Gg(v)$ and $\WinStrat_\Gg(v)$ to denote the set of all (winning) strategies of Player~0 from position $v \in \Gg$, and we drop $\Gg$ if the game is clear from the context.
\end{defi}

\begin{figure}
\newcommand{\enode}[4][]{\draw [arr] (#2) edge node [#1] {$#4$} (#3);}
\begin{subfigure}[t]{.46\linewidth}
\centering
\begin{tikzpicture}[node distance=1.7cm,baseline]
\path [draw=none] (0,1.1cm) rectangle (0,-3.2cm); % alignment hack
\node [p1,label={left:$v$}] (0) {};
\node [p1, right of=0, yshift=.7cm] (1) {};
\node [p0, F, right of=1, yshift=-.7cm, label={below:$v'$}] (2) {};
\node [p0, right of=2, yshift=.7cm] (3) {};
\node [p1, F, right of=2, yshift=-.7cm, label={below:$u$}] (4) {};
\node [p0, F, below of=2,label={above:$w$},yshift=-.2cm] (5) {};
\node [p1, left of=5] (6) {};
\draw [arr]
    (0) edge node {$a$} (1) (0) edge node {$c$} (2) (1) edge node {$b$} (2)
    (2) edge [bend left=20pt] node {$d$} (0)
    (2) edge node {$e$} (3)
    (3) edge [loop right] node {$i$} (3)
    (2) edge node [below left] {$f$} (4)
    (3) edge node {$h$} (4)
    (4) edge [loop right] node {$g$} (4)
    (4) edge node {$k$} (5)
    (5) edge [loop below] node {\strut$m$} (5)
    (5) edge node {$n$} (6)
    (6) edge [loop below] node {\strut$p$} (6)
    (6) edge [bend left] node {$q$} (0)
    ;
\end{tikzpicture}
\caption{Rectangular nodes belong to Pl.~1, round nodes to Pl.~0, dashed nodes are in $F$.}
\label{fig:RunningGame}
\end{subfigure}
\hfill
\begin{subfigure}[t]{.45\linewidth}
\centering
\begin{tikzpicture}[node distance=1.3cm,every node/.style={scale=0.75},baseline]
\path [draw=none] (0,1.1cm) rectangle (0,-3.2cm); % alignment hack
\node [p1,label={left:$v$}] (0) {};
\node [p1,right of=0,yshift=.8cm] (1) {};
\node [p0,F,right of=1,label={below:$v'$}] (2) {};
\node [p0,right of=2] (3) {};
\node [p1,F,right of=3] (4) {};
\node [marker,right of=4] (4a) {};
\node [marker,below of=4,xshift=.5cm] (4b) {};
\node [p0,F,right of=0,yshift=-.8cm,label={below:$v'$}] (a) {};
\node [p1,F,right of=a] (b) {};
\node [p1,F,below of=b,xshift=.5cm] (c) {};
\node [p1,F,below of=c,xshift=.5cm] (d) {};
\node [marker,right of=d] (d1) {};
\node [marker,below of=d,xshift=.5cm] (d2) {};
\node [p0,F,right of=b] (b1) {};
\node [p0,F,right of=b1] (b2) {};
\node [p0,F,right of=b2] (b3) {};
\node [marker,right of=b3] (b4) {};
\node [p0,F,right of=c] (c1) {};
\node [p0,F,right of=c1] (c2) {};
\node [p0,F,right of=c2] (c3) {};
\node [marker,right of=c3] (c4) {};
\enode 0 1 a
\enode 1 2 b
\enode 2 3 e
\enode 3 4 h
\enode[swap] 0 a c
\enode a b f
\enode[left] b c g
\enode[left] c d g
\enode b {b1} k
\enode {b1} {b2} m
\enode {b2} {b3} m
\enode c {c1} k
\enode {c1} {c2} m
\enode {c2} {c3} m
%
% draw dashed edges, clip for proper bounding box
\draw [densely dotted,shorten <=3pt,shorten >=10pt] (4) edge (4a) (4) edge (4b);
\begin{scope}
\draw [densely dotted,shorten <=3pt,shorten >=10pt] (b3) edge (b4);
\clip (c3.north west) rectangle ($(c3.south east)+(.4cm,-.4cm)$);
\draw [densely dotted,shorten <=3pt,shorten >=10pt] (c3) edge (c4);
\end{scope}
\begin{scope}
\clip (d.north west) rectangle ($(d.south east)+(.4cm,-.4cm)$);
\draw [densely dotted,shorten <=3pt,shorten >=10pt] (d) edge (d1) (d) edge (d2);
\end{scope}
\end{tikzpicture}
\caption{Depiction of an infinite strategy tree of a winning strategy for Pl.~0 from position $v$.}
\label{fig:RunningStrategy}
\end{subfigure}
\caption{Running example of a Büchi game and a winning strategy.}
\label{fig:Running}
\end{figure}

\begin{example}
An example of a Büchi game is depicted in \cref{fig:Running}.
Player~0 has essentially three different positional winning strategies from $v$,
by either choosing edge $d$, or edges $e,h,m$ or $f,m$.
Notice that for the first strategy, we did not specify moves for all positions in $V_0$ as these positions cannot be reached when edge $d$ is played; this is the main reason why we represent strategies as trees.
\Cref{fig:RunningStrategy} depicts such a tree representation of a strategy.
This strategy is a typical example of a winning strategy that is not positional, but still minimal if we take edge multiplicities (see \cref{defEdgeProfile}) into account.
\end{example}

\section{Strategies with Minimal Effort}

% samepage: we do not want to have a "lonely quote" right before the page break
\begin{samepage}
\begin{aquote}{Antoine de Saint-Exupéry}
La perfection est atteinte, non pas lorsqu’il n’y a plus rien à ajouter, mais lorsqu’il n’y a plus rien 
à retirer.\footnote{Perfection is achieved, not when there is nothing more to add, but when there is nothing left to take away.}
\end{aquote}

\noindent
As a measure for the complexity or effort of a strategy,
we consider the set of edges a strategy $\Ss$ uses and how often each of these edges appears in the strategy tree.
Under this measure, the simplest strategies are the ones that do not play redundant edges
-- hence no moves are left to take away.
\end{samepage}

\begin{defi}\label{defEdgeProfile}
Given an edge $e = vw \in E$ in a Büchi game $\Gg$ and a strategy $\Ss$ in $\Gg$,
we denote by $\ecount \Ss e = | \{ \rho v \in \Ss \mid \rho v \to \rho v w \text{ is an edge in } \Ss \}| \in \N \cup \{ \infty \}$ the number of times (possibly infinite) the edge $e$ occurs in $\Ss$.
With each strategy $\Ss$ we associate its \emph{edge profile}, the vector $\ep \Ss = (\ecount \Ss e)_{e \in E}$.
\end{defi}

\begin{example}\label{ex:redundantMove}
Consider the following Büchi game:
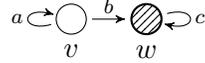
\begin{center}
\begin{tikzpicture}[baseline]
\node [p0,label={below:$v$},anchor=base,yshift=.1cm] (0) {};
\node [p0,F,label={below:$w$},right of=0] (1) {};
\draw [arr]
    (0) edge [loop left] node {$a$} (0)
    (1) edge [loop right] node {$c$} (1)
    (0) edge node {$b$} (1);
\end{tikzpicture}
\end{center}
Player~0 wins by first looping $n$ times at position $v$ (for any fixed $n \in \N$) and then moving to $w$, corresponding to the edge profile $(n,1,\infty)$.
Clearly, looping at $v$ is a redundant move, so we regard the strategy with $n=0$ as the simplest one (that wins with the least effort).
\end{example}

To formalize the intuition of redundant moves, we define an order $\absorb$ on strategies called \textit{absorption}.
This is defined in such a way that the $\absorb$-maximal strategies are the simplest ones
that avoid redundant moves whenever possible.

\begin{defi}
Let $\Ss_1,\Ss_2$ be two strategies in a Büchi game $\Gg = (V,V_0,V_1,E,F)$.
We say that $\Ss_1$ \emph{absorbs} $\Ss_2$, denoted $\Ss_1 \absorb \Ss_2$,
if $\ecount {\Ss_1} e \le \ecount {\Ss_2} e$ for all edges $e \in E$.
If additionally $\ecount {\Ss_1} e < \ecount {\Ss_2} e$ for some $e \in E$,
we say that $\Ss_1$ \emph{strictly absorbs} $\Ss_2$, denoted $\Ss_1 \absorbneq \Ss_2$.
They are \emph{absorption-equivalent}, denoted $\Ss_1 \equiv \Ss_2$, if both $\Ss_1 \absorb \Ss_2$ and $\Ss_2 \absorb \Ss_1$.

A strategy $\Ss \in \Strat(v)$ is \emph{absorption-dominant from position $v$}, if there is no strategy $\Ss' \in \Strat(v)$ with $\Ss' \absorbneq \Ss$.
It is further \emph{strictly} absorption-dominant, if there is no other strategy $\Ss' \in \Strat(v)$ with $\Ss' \absorb \Ss$, so no other strategy is absorption-equivalent to $\Ss$.
\end{defi}

Notice that absorption is simply the inverse pointwise order on the edge profiles.
In particular, $\Ss_1 \equiv \Ss_2$ if, and only if, $\ep {\Ss_1} = \ep {\Ss_2}$.
We next aim at understanding the relation between
(strictly) absorption-dominant strategies and the standard notion of positional strategies.
As a starter, we show that absorption-dominant strategies are not necessarily positional
(cf.\ \cite{GraedelTan20} for a similar example).

\begin{example}\label{ex:Weakpos}
Consider the strategy $\Ss$ as depicted in \cref{fig:RunningStrategy}.
It is not positional, as the choice for position $v'$ is not unique (both $e$ and $f$ occur in $\Ss$).
It is, however, absorption-dominant.
As there are two paths to $v'$, every strategy must either use $e$ or $f$ twice, or use both edges.
If $e$ (or $f$) is used twice, then the strategy cannot absorb $\Ss$, and one can verify that $\Ss$ absorbs all strategies using both $e$ and $f$.

It is not strictly absorption-dominant, as we obtain an absorption-equivalent strategy by switching the two branches in the depiction of $\Ss$, so that $e$ is used after $c$, and $f$ after $b$.
\end{example}

Strategies such as the one in \cref{fig:RunningStrategy} are not positional, but satisfy the weaker property that within each \emph{play}, the strategy makes a unique decision for each position $v \in V_0$.
This notion of strategies has been introduced as \emph{persistent} strategies in \cite{MarcinkowskiTruderung02} in the context of LTL on game graphs and has been further studied in \cite{Duparc03}.
Persistent strategies have also been called \emph{weakly positional} in \cite{GraedelTan20}.

We say that a strategy \emph{plays positionally} from a position $v \in V_0$ if the strategy makes a unique choice at position $v$ (not depending on the history of the play).
A strategy that plays positionally from all positions in $V_0$ is positional.
With this notation, we now clarify the relation between the different notions of strategies; a summary is shown in \cref{figStrategyClasses}.
We first observe that if a strategy $\Ss$ does not play positionally from $v$, we can always obtain a strategy $\Ss'$ with $\Ss' \absorb \Ss$ by swapping the choices at $v$, which leads to:

\begin{figure}[t]
\centering
\begin{tikzpicture}[font=\small, every label/.style={font=\scriptsize}, xscale=0.8,yscale=0.9]
\draw [pattern=crosshatch dots, pattern color=lightgray!30]  (0,.4) ellipse (4cm and 1.15cm);
\node[anchor=north] at (0,1.4) {absorption-dominant};
\draw [pattern=north east lines, pattern color=lightgray!60] (0,0) ellipse (2.6cm and .7cm) node [align=center,yshift=0pt] {positional \\[-.3em] = \\[-.2em] strictly abs.-dom.} ;
\draw (0,.8) ellipse (6cm and 1.6cm);
\node[anchor=north] at (0,2.3) {persistent};
\node [dot,label={right:Ex.~\ref{ex:Weakpos}}] at (2.2,0.7) {};
\node [dot,label={right:Ex.~\ref{ex:WeakposNotDominant}}] at (4.2,1.0) {};
\end{tikzpicture}
\caption{Venn diagram depicting various classes of winning strategies.}
\label{figStrategyClasses}
\end{figure}

\begin{prop}
Strictly absorption-dominant strategies coincide with positional strategies.
\end{prop}
\begin{proof}
Let $\Ss \in \Strat_\Gg(v)$ be a strategy from $v$.
First assume towards a contradiction that $\Ss$ is positional but not strictly absorption-dominant.
That is, there is a different strategy $\Ss' \in \Strat(v)$ with $\Ss' \absorb \Ss$.
Since $\Ss'$ is different from $\Ss$, there is a position $w$ and a path $\rho w$ occurring in both strategies
for which the strategies differ, i.e., we have $w_1 = \Ss(\rho w)$ and $w_2 = \Ss'(\rho w)$ with $w_1 \neq w_2$.
Since $\Ss$ is positional, the edge $ww_2$ does not occur in $\Ss$.
Hence it occurs strictly more often in $\Ss'$, contradicting the assumption $\Ss' \absorb \Ss$.

We prove the other direction by contraposition.
Let $\Ss$ be non-positional, so there is a position $w$ and two paths $\rho w$ and $\rho' w$ such that $\Ss(\rho w) \neq \Ss(\rho' w)$.
Let $\Ss_{\rho w}$ and $\Ss_{\rho' w}$ be the substrategies of $\Ss$ from $\rho w$ and $\rho' w$, respectively.
First assume that $\rho \prefix \rho'$.
We then consider the strategy $\Ss'$ that behaves like $\Ss$, but switches to $\Ss_{\rho' w}$ at $\rho w$.
As every edge occurring in $\Ss'$ also occurs in $\Ss$, we have $\Ss' \absorb \Ss$ and $\Ss$ is not strictly absorption-dominant.
The case $\rho' \prefix \rho$ is symmetric.
If, on the other hand, $\rho$ and $\rho'$ are incomparable nodes in $\Ss$,
we consider the strategy $\Ss'$ that behaves like $\Ss$, but plays $\Ss_{\rho' w}$ from $\rho w$ and $\Ss_{\rho w}$ from $\rho' w$, swapping the two substrategies.
Then $\Ss' \equiv \Ss$, so $\Ss$ is not strictly absorption-dominant.
\end{proof}

We next establish the relation to persistent strategies.
To this end, we first show under which circumstances absorption-dominant strategies must make unique choices.
Our proof needs the following combinatorial observation.

\begin{lem}\label{stratDominantFiniteEquiv}
Let $v \in \Gg$ be a position.
There are only finitely many absorption-dominant strategies from $v$ up to absorption-equivalence.
\end{lem}
\begin{proof}
Consider the pointwise order on edge profiles induced by the standard order on $\N \cup \{\infty\}$.
By definition, a strategy $\Ss$ is absorption-dominant from $v$ if, and only if, its edge profile $\ep \Ss$ is minimal among all strategies from $v$ (and absorption-equivalent strategies have the same edge profile).
By a simple combinatorial fact known as Dickson's lemma, every set of edge profiles contains only finitely many minimal elements.
\end{proof}

\begin{prop}\label{stratInfinitePositional}
Let $\Ss \in \WinStrat_\Gg(v)$ be absorption-dominant from $v$, and let $w \in V_0$ be a position.
If $w$ occurs infinitely often in $\Ss$, then $\Ss$ plays positionally from $w$.
\end{prop}

\begin{proof}
Consider the infinitely many substrategies at occurrences of $w$ in $\Ss$.
By \cref{stratDominantFiniteEquiv}, there is one such substrategy $\Ss_w$ such that infinitely many of the substrategies are absorption-equivalent to $\Ss_w$.
This means that every edge occurring in $\Ss_w$ also occurs in infinitely many substrategies and hence infinitely often in the full strategy $\Ss$.
Notice that $\Ss_w$ is winning from $w$, as it is a substrategy of the winning strategy $\Ss$.
Consider the subgame of $\Gg$ containing only edges occurring in $\Ss_w$.
Clearly, Player~0 wins from $w$ (using $\Ss_w$) and by positional determinacy, there is thus a positional winning strategy $\Ss_{\text{pos}}$ from $w$ using only edges that occur in $\Ss_w$ and hence infinitely often in $\Ss$.

Now consider the strategy $\Ss' \in \WinStrat_\Gg(v)$ that behaves like $\Ss$, but always uses $\Ss_{\text{pos}}$ from $w$. Then $\Ss' \absorb \Ss$ by construction of $\Ss_{\text{pos}}$.
Further, $\Ss_{\text{pos}}$ is positional and makes a unique choice $\Ss_{\text{pos}}(w)$.
If $\Ss$ would not play positionally from $w$, then there would be some path $\rho w$ such that $\Ss(\rho w) = w' \neq \Ss_{\text{pos}}(w)$.
But then the edge $w w'$ never occurs in $\Ss'$, so $\Ss' \absorbneq \Ss$ and $\Ss$ would not be absorption-dominant.
\end{proof}

With this important insight, we can deduce that the absorption-dominant winning strategies (from some position $v$) are a (strict) subset of the persistent strategies:
An absorption-dominant strategy must play positionally from positions that occur infinitely often; repetitions of positions that occur finitely often are always redundant.

\begin{cor}\label{stratDominantWeakpos}
Every absorption-dominant winning strategy in $\Gg$ is persistent.
\end{cor}

\begin{proof}
Let $\Ss \in \WinStrat_\Gg(v)$ be absorption-dominant from $v$.
Assume towards a contradiction that $\Ss$ is not persistent, so there is a position $w$ and a play of the form $\rho_1 w \rho_2 w \rho_3$ such that $\Ss$ makes different decisions at $w$, say $\Ss(\rho_1 w) = w_1$ and $\Ss(\rho_1 w \rho_2 w) = w_2$ with $w_1 \neq w_2$.
By \cref{stratInfinitePositional}, $w$ can only occur finitely often in $\Ss$.
Hence the edge $w w_1$ also occurs finitely often, say $n$ times.
Let $\Ss'_w$ be the substrategy of $\Ss$ from $\rho_1 w \rho_2 w$.
Now consider the strategy $\Ss' \in \WinStrat_\Gg(v)$ that behaves like $\Ss$, but switches to $\Ss'_w$ at $\rho_1 w$.
By construction, $\Ss'$ uses each edge at most as often as $\Ss$.
Moreover, one occurrence of the edge $w w_1$ is removed, so this edge occurs at most $n-1$ times in $\Ss'$.
Hence $\Ss' \absorbneq \Ss$, contradicting the absorption-dominance of $\Ss$.
\end{proof}

\begin{example}\label{ex:WeakposNotDominant}
For strictness, consider the following game (a modified part of \cref{fig:RunningGame}):

\begin{center}
\begin{tikzpicture}[scale=0.9,yscale=0.7,xscale=0.6,baseline={(0,0)}]
\node [p1] at (0,0) (1) {$v$};
\node [p1] at (1.5,1) (2a) {};
\node [p0,F] at (3,0) (3) {};
\node [p0,F] at (4.5,1) (4a) {};
\path [arr] (1) edge node {} (2a) (1) edge node [below left] {} (3)
      (2a) edge node {} (3)
      (3) edge node {$a$} (4a)
      (3) edge [in=-60,out=-20,min distance=1cm] node {$b$} (3)
      (4a) edge [loop right,min distance=.9cm] node {} (4a)
      ;
\end{tikzpicture}
\hspace{.5cm}
\parbox[c]{3.5cm}{
$\ecount {\Ss_1} {a,b} = (2,0)$, \\
$\ecount {\Ss_2} {a,b} = (0,\infty)$, \\
$\ecount {\Ss_3} {a,b} = (1,\infty)$.
}
\end{center}

\noindent
Due to the self-loop $b$, only the positional strategies $\Ss_1$ (always take $a$) and $\Ss_2$ (always take $b$) are absorption-dominant from $v$.
The strategy $\Ss_3$ that, depending on Player~1's choice, either takes edge $a$ or loops indefinitely using edge $b$ is persistent, but not absorption-dominant:
it is strictly absorbed by $\Ss_2$.
\end{example}

As a consequence of \cref{stratDominantWeakpos}, all moves after the first repeated position are determined by persistence.
We can thus represent absorption-dominant strategies in a compact way and strengthen \cref{stratDominantFiniteEquiv} as follows.

\begin{cor}
Let $\Gg$ be a game with $n = |V|$ positions.
Every winning strategy $\Ss \in \WinStrat_\Gg(v)$ that is absorption-dominant from $v$
can be uniquely represented by a subtree of the tree unraveling of height at most $n$.
In particular, the number of absorption-dominant winning strategies is finite.
\end{cor}

\section{A Whirlwind Tour of Semiring Semantics}

This section gives an overview on semiring semantics for fixed-point logics,
with a focus on the semirings relevant for the case study.
For a complete account, we refer to \cite{DannertGraNaaTan21}.

\subsection{Semirings}\label{sect:semirings}

Semirings are algebraic structures with two binary operations, usually denoted $+$ and $\bcdot$,
which we use to interpret the logical connectives $\lor$ and $\land$.
While semirings are very general structures, we make additional assumptions to ensure well-defined and meaningful semiring semantics for logics with fixed-point operators.

\begin{defi}
A \emph{commutative semiring} is an algebraic structure 
$(\K,+,\bcdot,0,1)$, with $0\neq1$,  such that $(\K,+,0)$
and $(\K,\bcdot,1)$ are commutative monoids, $\bcdot$
distributes over $+$, and $0\bcdot a=a\bcdot 0=0$.
It is \emph{idempotent} if $a+a = a$ for all $a \in \K$.
\end{defi}

All semirings we consider are commutative, so we omit \emph{commutative} in the following.
Towards fixed-point logic, we compute least and greatest fixed points with respect to the natural order $\le_\K$ (see below) and to ensure that they exist, we require $\le_\K$ to be a complete lattice (in fact, suprema and infima of chains would suffice, but in idempotent semirings this is equivalent).
We additionally impose a natural \emph{continuity} requirement which is crucial to our proofs, but does not seem to be a strong restriction in practice (we are not aware of any natural complete-lattice semirings that are not continuous).
Regarding notation, a \emph{chain} is a totally ordered set $C \subseteq \K$ and we write
$a \circ C = \{ a \circ c \mid c \in C\}$ for $a \in \K$.

\begin{defi}
In an idempotent semiring $(\K,+,\bcdot,0,1)$, the \emph{natural order} $\le_\K$ is the partial order defined by $a \le_\K b \Leftrightarrow a+b=b$.
We say that $\K$ is \emph{fully continuous} if $\le_\K$ is a complete lattice (with supremum $\Sup$ and infimum $\Inf$) and for all non-empty chains $C \subseteq \K$, elements $a \in \K$ and $\circ \in \{+,\bcdot\}$,
\[
    \Sup (a \circ C) = a \circ \Sup C,
    \quad \text{and} \quad
    \Inf (a \circ C) = a \circ \Inf C.
\]
A semiring homomorphism $h \colon \K_1 \to \K_2$ on fully-continuous semirings is \emph{fully continuous} if $h(\Sup C) = \Sup h(C)$ and $h(\Inf C) = \Inf h(C)$ for all non-empty chains $C \subseteq \K_1$.
\end{defi}

By the Knaster-Tarski theorem, every $\le_\K$-monotone function $f \from \K \to \K$ on a fully-continuous semiring has a least fixed point $\lfp(f)$ and a greatest fixed point $\gfp(f)$ in $\K$,
and this suffices to guarantee well-defined semiring semantics of fixed-point logics.
However, from a provenance perspective we further want this semantics to be meaningful in the sense that the value of a formula provides insights into why the formula holds.
It turns out that this is the case if we additionally require the semiring to be absorptive (see \cite{DannertGraNaaTan21}).

\begin{defi}
A semiring $\K$ is \emph{absorptive} if $a+ab = a$ for all $a,b \in \K$.
\end{defi}

We remark that absorption is equivalent to $\K$ being \emph{0-closed} or \emph{bounded} \cite{Mohri02},
that is, $1+a = 1$.
If $\K$ is idempotent, then absorption is further equivalent to multiplication being decreasing,
that is, $a \bcdot b \le_\K a,b$.
Clearly, every absorptive semiring is idempotent and thus partially ordered by $\le_\K$, with $1$ as top element.
If we additionally assume full continuity, we can extend any absorptive semiring by an infinitary power operation
$a^\infty = \Inf_{n \in \N} a^n$ with natural properties such as $a \bcdot a^\infty = a^\infty$, $(ab)^\infty = a^\infty b^\infty$ and $(a+b)^\infty = a^\infty + b^\infty$.

\begin{example}
Here is a short, non-exhaustive list of semirings used in provenance analysis of databases and logics \cite{GreenKarTan07,GreenTan17,GraedelTan20}.
\begin{itemize}
\item The \emph{Boolean semiring} $\Bool=(\{\Bzero,\Bone\},\vee,\wedge,\Bzero,\Bone)$ is the standard habitat of 
logical truth. It is absorptive and (trivially) fully continuous.
\item $\Nat=(\Nat,+,\cdot,0,1)$ is used for counting evaluation strategies for a logical statement.
It is not absorptive and hence not well suited for fixed-point logics.
\item The \emph{Viterbi} semiring $\Vit=([0,1],\max,\cdot,0,1)$
is used to compute \emph{confidence scores}
of logical statements. It is isomorphic to the \emph{tropical} semiring $\Trop=(\mathbb{R}_{+}^{\infty},\min,+,\infty,0)$ 
used for measuring the cost of evaluation strategies.
Both are absorptive and fully continuous.

\item The \emph{min-max} semiring $(A, \max, \min, a, b)$ on a totally ordered set $(A,\leq)$
with least and greatest elements $a$ and $b$ can be used to model access privileges. It is absorptive and fully continuous. \cornerhere
\end{itemize}
\end{example}

From now on, all semirings we consider are commutative, absorptive and fully continuous.
Besides the application semirings listed above, we are particularly interested in universal semirings of polynomials to represent abstract information.
We can then use fully-continuous homomorphisms to specialize the computed information to application semirings as needed, as these homomorphisms preserve fixed points.
The common examples of semirings of polynomials $\N[X]$ and formal power series $\N^\infty \ps X$,
as used for provenance analysis of FO and Datalog in \cite{GreenKarTan07, GraedelTan17}, are not absorptive and hence not well-suited for fixed-point logic.

Instead, we rely on semirings of (generalized\footnote{The definition we use here generalizes the one in \cite{DeutchMilRoyTan14} by allowing $\infty$ as exponent.}) \emph{absorptive polynomials} (cf.\ \cite{DannertGraNaaTan21}).
Essentially, an absorptive polynomial such as $ab^3 + c^\infty$ is a sum of monomials over a finite set of variables $X$,
but without coefficients and with exponents from $\N \cup \{\infty\}$.
Monomial multiplication is defined as usual by adding exponents (with $n+\infty = \infty$).
The key ingredient is absorption among monomials:
we say that a monomial $m_1$ \emph{absorbs} $m_2$ ($m_1 \absorb m_2$), if all its exponents are smaller (or equal).
Formally, $m_1 \absorb m_2$ if $m_1(x) \le m_2(x)$ for all $x \in X$, where $m_1(x)$ denotes the exponent of $x$ in $m_1$.
For example, $ab^2 \absorb a^\infty b^2$ and $a \absorb ab$, but $a^2b$ and $ab^2$ are incomparable.
In an absorptive polynomial, we omit all monomials that would be absorbed, so absorptive polynomials
are precisely the $\absorb$-\emph{antichains} of monomials. Consequently, addition and multiplication
are defined as usual, but afterwards we only keep the $\absorb$-maximal monomials.
For example, $(ab^2 + a^2b) \bcdot a^\infty = a^\infty b^2 + a^\infty b = a^\infty b$.

\begin{defi}
We write $\Sinf[X]$ for the semiring of absorptive polynomials over the finite variable set $X$.
Its elements are the $\absorb$-antichains of monomials (written as sums) with the operations described above.
The neutral elements are the empty polynomial (denoted by $0$) and the single monomial $1$ (with all zero exponents).
\end{defi}

This defines an absorptive, fully-continuous semiring \cite{DannertGraNaaTan21}.
In fact, $\Sinf[X]$ is the most general such semiring:

\begin{thm}[Universal property, \cite{DannertGraNaaTan21}]
\label{universality}
Every mapping $h \colon X \to \K$ into an absorptive, fully-continuous semiring $\K$
uniquely extends to a fully-continuous semiring homomorphism $h \colon \Sinf[X] \to \K$
(by means of polynomial evaluation).
\end{thm}

\subsection{Logic}

We consider here the fixed-point logic LFP that extends first-order logic FO by least and greatest fixed-point formulae of the form $\psi(\ty) = [\lfp R \tx.\ \phi(R,\tx)](\ty)$ and $\psi(\ty) = [\gfp R \tx.\ \phi(R,\tx)](\ty)$. Here, $R$ is a relation symbol occurring only positively in $\phi$ and $\tx,\ty$ are variable tuples of matching arity.
Given a (Boolean) model $\AA$ and a tuple $\ta$ of elements of $\AA$,
the formula $\psi(\ta)$ holds in $\AA$, denoted $\AA \models \psi(\ta)$,
if $\ta$ is contained in the least (or greatest) fixed point of the operator
$F_\phi \colon R \mapsto \{ \ta \mid \AA \models \phi(R,\ta) \}$
that maps a relation $R$ to the relation consisting of those tuples for which $\phi$ holds.
For more background and a precise definition, we refer to \cite{Graedel+07}.

In order to generalize Boolean semantics to semiring semantics,
we first adapt the notion of a model $\AA$.
Instead of determining for each literal whether it is true or false in $\AA$,
we assign to each literal a semiring value, interpreting $0$ as \emph{false} and all other values as \emph{nuances of true}.
Special care is required to ensure that the assignment is consistent with respect to opposing literals (this is not always necessary, but often desirable).
In the following, let $\K$ be a semiring, $A$ a finite universe and $\tau$ a relational signature
(we drop $A$ and $\tau$ if clear from the context).
We denote the set of (instantiated) literals as
\begin{align*}
    \Lit_{A,\tau} ={} &\{ R \ta, \neg R \ta \mid R \in \tau \text{ of arity $k$}, \ta \in A^k \}
    \,\cup\, \{\ta = \tb, \ta \neq \tb \mid \ta,\tb \in A^k\}.
\end{align*}
Given a literal $\lit$, we write $\neg \lit$ for the opposing literal (identifying $\neg \neg \lit$ and $\lit$).
The role of the Boolean model $\AA$ is then replaced by a semiring interpretation $\pi$ that assigns semiring values to all literals.

\begin{defi}\label{defKInterpretation}
Let $\K$ be a semiring. A \emph{$\K$-interpretation} (over finite $A$ and $\tau$) is an assignment $\pi\colon \Lit_{A,\tau} \to \K$ that maps true (in)equalities to $1$ and false (in)equalities to $0$.
We say that $\pi$ is \emph{model-defining}, if for each literal $\lit$, exactly one of $\pi(\lit)$ and $\pi(\neg \lit)$ is $0$.
\end{defi}

For Büchi games, we will always use the signature $\tau = \{ E, F, V_0, V_1 \}$, where $E$ is a binary and $F,V_0,V_1$ are unary relation symbols.
We can then view a game $\Gg = (V, V_0, V_1, E, F)$ as a $\tau$-structure.
Notice that we do not distinguish between the edge relation $E$ of $\Gg$ and the relation symbol $E$; it will always be clear from the context what we refer to.
The set of instantiated literals $\Lit_{V,\tau}$ then contains, e.g., $Ev_1 v_2$ and $\neg F v_1$, where $v_1,v_2 \in \Gg$.

We lift $K$-interpretations $\pi$ from literals to LFP-formulae in negation normal form ($\nnf$), resulting in a semiring value $\pi \ext \psi$, by interpreting the logical connectives as semiring operations.
For fixed-point formulae, we consider the induced operator $F_\phi$ analogous to the Boolean case, but acting on functions $g \from A^k \to \K$ instead of relations $R \subseteq A^k$ (which can be seen as functions $R \from A^k \to \Bool$, justifying our generalisation).
We extend the natural order to such functions by pointwise comparison.
More formally, given a $\K$-interpretation $\pi$ over signature $\tau$, we denote by $\pi[R\mapsto g]$ the $\K$-interpretation over $\tau \cup \{R\}$ obtained from $\pi$ by adding values $g(\tup a)$ for the instantiated atoms $R\tup a$.
The analogue of the Boolean operator $F_\phi$ is then the operator $F_\pi^\phi$ that maps a function $g \colon A^k \to \K$ to the function
\[   F_\pi^\phi(g) \colon \; \tup a\mapsto   \pi[R\mapsto g]\ext{\phi(R,\tup a)}. \]
With this in mind, we define the following natural generalization of Boolean semantics.

\begin{defi}
 A $\K$-interpretation $\pi\colon \Lit_A(\tau)\rightarrow \K$
 (for finite $A$ and $\tau$)
 in a fully-continuous semiring $\K$
 extends to a $\K$-\emph{valuation}
 $\pi\colon \LFP(\tau)\rightarrow \K$ by mapping an $\LFP$-sentence $\psi(\tup{a})$ in negation normal form to a value $\pi\ext\psi$ using the rules
 \begin{align*}
  \pi\ext{\psi\vee\phi}&\coloneqq\pi\ext\psi + \pi\ext\phi, &
  \pi\ext{\exists x\psi(x)}&\coloneqq\textstyle\sum_{a\in A}\pi\ext{\phi(a)}, &
  \pi\ext{\neg\psi}&\coloneqq\pi\ext{\nnf(\neg\psi)}, \\
  \pi\ext{\psi\wedge\phi}&\coloneqq\pi\ext\psi \bcdot \pi\ext\phi, &
  \pi\ext{\forall x\psi(x)}&\coloneqq\textstyle\prod_{a\in A}\pi\ext{\phi(a)},
 \end{align*}
 and, for fixed-point formulae,
 \begin{align*}
  \pi \ext{[\lfp R\tup x. \phi(R,\tup x)](\tup a)} &\coloneqq \lfp(F_\pi^\phi)(\tup a), &
  \pi \ext{[\gfp R\tup x. \phi(R,\tup x)](\tup a)} &\coloneqq \gfp(F_\pi^\phi)(\tup a).
 \end{align*}
\end{defi}

An important property of the resulting semantics is that it is preserved by fully-continuous semiring homomorphisms,
in particular by polynomial evaluation in $\Sinf[X]$ due to \cref{universality} (but not by polynomial evaluation of $\Nat[X]$ or formal power series!).

\begin{thm}[Fundamental property, \cite{DannertGraNaaTan21}]
\label{fundamental}
Let $h \colon \K_1 \to \K_2$ be a fully-continuous semiring homomorphism.
Then for every $\K_1$-interpretation $\pi$, the mapping $h \circ \pi$ is a $\K_2$-interpretation and $h(\pi \ext \phi) = (h \circ \pi) \ext \phi$, for every $\phi\in\LFP$.
\end{thm}

\begin{center}
\begin{tikzpicture}[baseline,node distance=1.7cm,font=\small]
\node [baseline, anchor=base] (lit) {$\Lit_A(\tau)$};
\node [below left of=lit] (litS) {$K_1$};
\node [below right of=lit] (litT) {$K_2$};

\node [right=5cm of lit.base, anchor=base] (fol) {\upshape LFP};
\node [below left of=fol] (folS) {$K_1$};
\node [below right of=fol] (folT) {$K_2$};

\node [align=center] at ({$(lit.center)!0.5!(fol.center)$} |- {$(lit.center)!0.5!(litS.center)$}) {$\implies$};

\path[draw,->,shorten <=1pt, shorten >=1pt, font=\scriptsize]
(lit) edge node [above left] {$\pi$} (litS)
(lit) edge node [above right] {$h \circ \pi$} (litT)
(litS) edge node [above] {$h$} (litT)
(fol) edge node [above left] {$\pi$} (folS)
(fol) edge node [above right] {$h \circ \pi$} (folT)
(folS) edge node [above] {$h$} (folT)
;
\end{tikzpicture}
\end{center}

\section{Computing Strategies with Semiring Semantics}
\label{sec:TheFormula}

This section connects the previous sections on semiring semantics and absorption-dominant strategies.
We focus on the formula for the winning region in a Büchi game and show that its value under semiring semantics can be understood in terms of (absorption-dominant) winning strategies.

\begin{figure*}
\centering
\begin{tabular}{r|ccc}
\toprule
Interpretation & $\pitrack$ & $\pirev$ & $\piwin$ \\
\& application & \textit{strategy tracking} & \textit{reverse analysis of moves} & \textit{target synthesis} \\  \midrule
Semiring & $\Sinf[X]$ & $\Sinf[X,\nnX]$ or $\PosBool[X,\nnX]$ & $\PosBool[X,\nnX]$ \\
$\pi(Evw)$ & $X_{vw}/0$ & $X_{vw}$ (or $1/0)$ & $1/0$ \\
$\pi(\neg Evw)$ & $1/0$ & $\nn{X_{vw}}$ (or $1/0$) & $1/0$ \\
$\pi(F v)$ & $1/0$ & $1/0$ & $X_v$ \\
$\pi(\neg F v)$ & $1/0$ & $1/0$ & $\nn{X_v}$  \\
other literals & $1/0$ & $1/0$ & $1/0$ \\
\bottomrule
\end{tabular}
\caption{Semiring interpretations used in this paper (the notation $a/0$ indicates the value $a$ if the literal is true,
and $0$ if it is false in $\Gg$).}
\label{fig:Interpretations}
\end{figure*}

\subsection{The Semiring Interpretation}

We want to use semiring semantics to analyze moves in winning strategies.
For this reason, we label edges with indeterminates $X$ (cf.\ \cref{fig:RunningGame}) and use an $\Sinf[X]$-interpretation $\pitrack$ to track moves (i.e., edge literals $Evw$ for positions $v,w$) via their indeterminates.
We assume the game graph to be fixed and do not wish to track information about the target set $F$ or the active player at a certain node, hence we simply map all other literals over $\tau = \{E,F,V_0,V_1\}$, such as $F v$, $V_0 v$ and $\neg Evw$, to $0$ or $1$, depending on whether they are true or false in the fixed game.
The resulting interpretation is almost Boolean and hence behaves very similar to the original game, except that we remember which edges are used in the evaluation of a formula.

\begin{defi}
Let $\Gg = (V, V_0, V_1, E, F)$ be a Büchi game and let $X = \{ X_{vw} \mid vw \in E \}$ be a set of indeterminates for all edges.
We define the $\Sinf[X]$-interpretation $\pitrack \from \Lit_{V,\tau} \to \Sinf[X]$ as follows (depending on $\Gg$):
\begin{align*}
    \pitrack(Evw) &= X_{vw} \text{ for all edges } vw \in E, \quad \\
    \pitrack(\lit) &= \begin{cases}
    1, &\text{ if } \Gg \models \lit, \\
    0, &\text{ if } \Gg \not\models \lit,
    \end{cases} \; \text{ for all other literals $\lit \in \Lit_{V,\tau}$.}
\end{align*}
\end{defi}

For the applications in \cref{sec:Applications}, we may consider other interpretations which are defined in a similar way, but also track negative edge literals or the target set $F$.
An overview is given in \cref{fig:Interpretations}.
In this section, we always work with $\pitrack$.

\subsection{The Formula}
\label{subsec:TheFormula}

It is well known that the winning region (of Player~0) in a Büchi game is definable in fixed-point logic.
Intuitively, the winning region is the largest set $Y$ such that from each position in $Y$,
Player~0 can enforce a visit to $Y \cap F$ (after at least one move).
In LFP, we can express the winning region as follows (see, e.g., \cite{CanavoiGraLesPak15,Walukiewicz02}):
\[
  \FormulaWin(x) \coloneqq{} \big[\gfp Y y.\ [\lfp Z z.\ \phi(Y,Z,z)](y) \big](x),
\]
where
\begin{align*}
  \phi(Y,Z,z) \coloneqq{} &\Big(Fz \;\land\; ((V_0 z \land \E u (Ezu \land Yu)) 
  \lor (V_1 z \land \A u (Ezu \to Yu)))\Big) \\
  {}\lor{} &\Big(\neg Fz \;\land\; ((V_0 z \land \E u (Ezu \land Zu))
  \lor (V_1 z \land \A u (Ezu \to Zu)))\Big).
\end{align*}

Given a $\K$-interpretation $\pi$ for a Büchi game $\Gg = (V,V_0,V_1,E,F)$, semiring semantics of the above formula induces%
\footnote{Here we first translate $Ezu \to Yu$ to the formula $\neg Ezu \lor (Ezu \land Yu)$ in negation normal form.}
the following fixed-point computation.
To simplify the presentation, we introduce two families of variables,
$\mathbf Y = (Y_v)_{v \in V}$ and $\mathbf Z = (Z_v)_{v \in V}$ that take values in $\K$.
We can then express the resulting semiring valuation as $\pi \ext {\FormulaWin(v)} = Y^*_v$ where
$\mathbf Y^* = (Y_v^*)_{v \in V}$ is the \emph{greatest} solution to the equation system
\[
    \mathbf Y = \mathbf Z^*(\mathbf Y)
\]
where, in turn, $\mathbf Z^*(\mathbf Y)$ is the \emph{least} solution, given values $\mathbf Y = (Y_v)_{v \in V}$, to the equation system consisting of the following equation for all $v \in V$:
\begin{align*}
    Z_v ={} &\pi(Fv) \bcdot \Big((\pi(V_0 v) \bcdot \sum_{w \in V} (\pi(Evw) \bcdot Y_w)) + (\pi(V_1 v) \bcdot \prod_{w \in V} (\pi(\neg Evw) + \pi(Evw) \bcdot Y_w))\Big) \\
    {}+{} &\pi(\neg Fv) \bcdot \Big((\pi(V_0 v) \bcdot \sum_{w \in V} (\pi(Evw) \bcdot Z_w)) + (\pi(V_1 v) \bcdot \prod_{w \in V} (\pi(\neg Evw) + \pi(Evw) \bcdot Z_w))\Big).
\end{align*}

For most of this paper, we use $\pitrack$ to track only \emph{moves} of winning strategies.
As $\pitrack$ maps most of the literals to $0$ or $1$, we can simplify the equations depending on $v$:

\begin{center}
\renewcommand{\arraystretch}{1.5}
\begin{tabular}{c|c|c}
& $v \in F$ & $v \notin F$ \\[-2pt] \hline
$v \in V_0$ &
    $\displaystyle Z_v = \sum_{w \in vE} \pi(Evw) \bcdot Y_w$ &
    $\displaystyle Z_v = \sum_{w \in vE} \pi(Evw) \bcdot Z_w$ \\
$v \in V_1$ &
    $\displaystyle Z_v = \prod_{w \in vE} \pi(Evw) \bcdot Y_w$ &
    $\displaystyle Z_v = \prod_{w \in vE} \pi(Evw) \bcdot Z_w$
\end{tabular}
\end{center}

A good way to think about (and compute) the least and greatest solutions is the fixed-point iteration.
The idea is to start with each $Z_v$ set to the least element $0$ of the semiring, then apply the above equations (i.e., the induced operator $F^\phi_{\pitrack}$) to compute a next, larger semiring value and repeat this process until a fixed-point is reached (notice that the iteration can also be infinite, the fixed-point is then the supremum or infimum).

\begin{example}\label{exComputation}
Recall the simple game from \cref{ex:redundantMove}
$\big($\!\!
\begin{tikzpicture}[baseline,font=\scriptsize]
\node [p0,label={below:$v$},anchor=base,yshift=.1cm] (0) {};
\node [p0,F,label={below:$w$},right of=0] (1) {};
\draw [arr]
    (0) edge [loop left] node {$a$} (0)
    (1) edge [loop right] node {$c$} (1)
    (0) edge node {$b$} (1);
\end{tikzpicture}
$\!\!\big)$.

\newcommand{\vv}[2]{\begin{pmatrix}#1 \\ #2\end{pmatrix}}
\newcommand{\tY}{\mathbf Y}
\newcommand{\tZ}{\mathbf Z}
\newcommand{\arrF}{\xmapsto{\!\!\!F_{\pitrack}^\phi\!}}

\noindent
Using the interpretation $\pitrack$ corresponding to the edge labels,
we obtain the following fixed-point iteration.
We write the tuples $\tY$ and $\tZ$ as vectors $({Y_v} \; {Y_w})^T$ and $({Z_v} \; {Z_w})^T$.
\begin{center}
\begin{tikzpicture}[font=\small,node distance=1.5cm]
\node (Y) {$\tY:$};
\node (Z) [below=.7cm of Y] {$\tZ:$};

\node [right=0cm of Y] (Y1) {$\vv 1 1$};
\node [xshift=.7cm] at (Z -| Y1) (Z11) {$\vv 0 0$};
\node [right of=Z11] (Z12) {$\vv 0 c$};
\node [right of=Z12] (Z13) {$\vv {bc} c$};

\node [xshift=.7cm] (Y2) at (Y1 -| Z13) {$\vv {bc} c$};
\node [xshift=.7cm] at (Z -| Y2) (Z21) {$\vv 0 0$};
\node [right of=Z21] (Z22) {$\vv 0 {c^2}$};
\node [right of=Z22] (Z23) {$\vv {bc^2} {c^2}$};

\node [xshift=.7cm] (Y3) at (Y1 -| Z23) {$\vv {bc^2} {c^2}$};
\node [xshift=.8cm] at (Z -| Y3) (Z31) {$\;\dots\vphantom{\vv x x}$};

\node [xshift=.9cm] (Y4) at (Y1 -| Z31) {$\vv {bc^n} {c^n}$};
\node [xshift=.8cm] at (Z -| Y4) (Z41) {$\;\dots\vphantom{\vv x x}$};

\path [draw,|->,every node/.style={anchor=base,yshift=4pt}]
    (Z11) edge node {$F^\phi$} (Z12)
    (Z12) edge node {$F^\phi$} (Z13)
    (Z21) edge node {$F^\phi$} (Z22)
    (Z22) edge node {$F^\phi$} (Z23)
;

\path [draw,->,short=-3pt]
    (Y1) edge (Z11)
    (Z13) edge (Y2)
    (Y2) edge (Z21)
    (Z23) edge (Y3)
    (Y3) edge (Z31)
    (Z31) edge (Y4)
    (Y4) edge (Z41)
;
\end{tikzpicture}
\end{center}

\noindent
We obtain the overall result $\pitrack \ext {\FormulaWin(v)} = Y_v^* = \Inf_n bc^n = bc^\infty$ corresponding to the unique absorption-dominant strategy using edge $b$ once and $c$ infinitely often (cf.\ \cref{ex:redundantMove}).
\end{example}

\subsection{Connection to Strategies}

By mapping edges to semiring values, we can track edges through the fixed-point computation.
In \cref{exComputation}, the resulting semiring value revealed how often each edge is used in the unique absorption-dominant winning strategy.
We now generalize this observation.
For simplicity, we only consider $\K$-interpretations $\pi$ that are \emph{edge tracking} for a given game $\Gg$.
That is, they may assign arbitrary values to positive edge literals $Evw$, but all other literals are mapped to $0$ or $1$ in accordance with $\Gg$.
To make the connection to strategies explicit, we first define semiring values for strategies based on the appearance of edges.

\begin{defi}
Let $\Ss$ be a strategy in a Büchi game $\Gg = (V, V_0, V_1, E, F)$.
Let $\K$ be an absorptive, fully-continuous semiring and $\pi$ an edge-tracking $\K$-interpretation on $\Gg$.
The \emph{$\K$-value} of $\Ss$ is the product of the values for all edges appearing in $\Ss$.
Formally,
\[
    \pi \ext \Ss \coloneqq \prod_{vw \in E} \pi(Evw)^{\ecount \Ss {vw}},
\]
where infinite exponents are interpreted by the infinitary power operation of the semiring.
\end{defi}

The semiring value of $\FormulaWin$ can then be expressed as the sum over the values of all winning strategies.
A direct proof is not completely straightforward, as fixed-point iterations and strategy trees can both be infinite (even if $\Gg$ is finite).
Instead, we make use of a similar sum-of-strategies result for model-checking games for LFP (see \cite{DannertGraNaaTan21}).

\begin{thm}[Sum of Strategies]
\label{thmSumOfStrategiesTracking}
Let $\Gg$ be a Büchi game and $v$ a position in $\Gg$.
Let $\K$ be an absorptive, fully-continuous semiring and $\pi$ an edge-tracking $\K$-interpretation.
Then,
\[
    \pi \ext {\FormulaWin(v)} = \sum \big\{ \pi \ext \Ss \;\big|\;
    \text{$\Ss \in \WinStrat_\Gg(v)$ is absorption-dominant from $v$} \big\}.
\]
\end{thm}

It is in fact this central result that motivated the notion of \emph{absorption-dominant} strategies.
However, as we have already discussed, these may also be interesting in their own right if one is interested in minimal winning strategies.

\begin{example}
For the edge-tracking interpretation $\pitrack$ induced by the edge labels in  \cref{fig:RunningGame}, we obtain
\begin{align*}
    \pitrack \ext {\FormulaWin(v)} ={}
    &(abcd)^\infty + 
    abc \, e^2 h^2 (gkm)^\infty +
    abc \, f^2 (gkm)^\infty +
    abc \, ef h (gkm)^\infty.
\end{align*}
There are four monomials, corresponding to four equivalence classes of absorption-dominant strategies.
Each monomial reveals the edges that appear in the corresponding strategies, so we see that the first three monomials belong to positional (and hence uniquely defined) strategies.
The last monomial belongs to the non-positional strategy shown in \cref{fig:RunningStrategy} (and its switched version, see \cref{ex:Weakpos}).
The values of all other strategies are strictly absorbed by one of these monomials.
\end{example}

\subsection{Proof of the Sum-of-Strategies Theorem}

% model checking games need tweaked tikz style
% use begingroup to make tikzset local to this subsection
\begingroup
\newcommand{\MC}[2]{\mathsf{MC}(#1,#2)} % \MC \Gg v
\newcommand{\MCP}[2]{\widetilde{\mathsf{MC}}(#1,#2)} % \MC \Gg v
\tikzset{
    arr/.style={draw,->,>=stealth',shorten <=2pt,shorten >=2pt,every node/.style={auto,inner sep=2pt,font=\scriptsize}},
    gamenode/.style={draw,inner sep=2pt,minimum size=.5cm,font=\small},
    lit/.style={gamenode,draw=none},
    p0/.style={gamenode,rectangle,rounded corners=.25cm},
    p1/.style={gamenode,rectangle},
    F/.style={thick, preaction={fill,pattern=north east lines,opacity=0.4},font=\boldmath\small},
    dot/.style={circle,draw,fill,black,minimum size=3pt,inner sep=0pt},
    marker/.style={draw=none,inner sep=0pt,overlay},
    short/.style={ shorten >=#1, shorten <=#1 }
}

\begin{figure*}
\centering
\begin{tikzpicture}[font=\small,node distance=.6cm]
% game for v1
  \node [p0,F] (Y1) {$Y v_1$};
  \node [p0,right=of Y1] (Z1) {$Zv_1$};
  \node [draw,cloud,minimum width=11cm,minimum height=3cm,cloud puffs=30,densely dotted,aspect=1,cloud puff arc=100,right=of Z1,text=gray] (C1) {};
% game for vn
  \node [p0,F,below=14cm of Y1] (Yn) {$Y v_n$};
  \node [p0,right=of Yn] (Zn) {$Zv_n$};
  \node [draw,cloud,minimum width=11cm,minimum height=3cm,cloud puffs=30,densely dotted,aspect=1,cloud puff arc=100,right=of Zn,text=gray] (Cn) {};
% main game
  \node [p0,F,below=7cm of Y1] (Y) {$Yv_i$};
  \node [p0,right=.4cm of Y] (Z) {$Zv_i$};
  \node [p0,right=.4cm of Z] (phi) {$\phi(v_i)$};
  \node [p1,right=.4cm of phi,yshift=1.8cm] (F) {$Fv_i \land \theta_1(v_i)$};
  \node [p0,right=of F] (t1) {$\theta_1(v_i)$};
  \node [p1,right=of t1,yshift=.7cm] (V0) {$V_0 v_i \land \dots$};
  \node [p1,right=of t1,yshift=-.7cm] (V1) {$V_1 v_i \land \dots$};
% existential quantifier
  \node [p0,right=of V0] (Ex) {$\E u \dots$};
  \node [p1,right=of Ex,yshift=1cm,minimum height=.2cm,align=center] (E1) {\dots};
  \node [p1,right=of Ex,yshift=.4cm,minimum height=.2cm,align=center] (E2) {\dots};
  \node [p1,right=of Ex,yshift=-.2cm,minimum height=.2cm,align=center] (E3) {\dots};
  \node [lit,right=of E1] (Ev1) {$E v_i v_1$};
  \node [lit,right=of E2] (Ev2) {$E v_i v_i$};
  \node [lit,right=of E3] (Ev3) {$E v_i v_n$};
% universal quantifier
  \node [p1,right=of V1] (All) {$\A u \dots$};  
  \node [p1,right=of All,yshift=.2cm,minimum height=.2cm,align=center] (A1) {\dots};
  \node [p1,right=of All,yshift=-.4cm,minimum height=.2cm,align=center] (A2) {\dots};
  \node [p1,right=of All,yshift=-1cm,minimum height=.2cm,align=center] (A3) {\dots};
  \node [lit,right=of A1] (Av1) {$E v_i v_1$};
  \node [lit,right=of A2] (Av2) {$E v_i v_i$};
  \node [lit,right=of A3] (Av3) {$E v_i v_n$};
% lower part (mostly omitted)
  \node [p1,right=.4cm of phi,yshift=-1.8cm] (nF) {$\neg Fv_i \land \theta_1(v_i)$};
  \node [p0] at (t1 |- nF) (t2) {$\theta_2(v_i)$};
  \node [draw,cloud,minimum width=4cm,cloud puffs=30,densely dotted,aspect=3.3,cloud puff arc=100,right=of t2,text=gray,align=center] (analog) {analogous to $\theta_1(v_i)$,\\but move to $Z v_j$ afterwards};
% remaining literals
  \node [lit,below=.5cm of F] (Fv) {$Fv_i$};
  \node [lit,above=.5cm of nF] (nFv) {$\neg Fv_i$};
  \node [lit,above=.5cm of V0] (V0v) {$V_0 v_i$};
  \node [lit,below=.5cm of V1] (V1v) {$V_1 v_i$};
% main arrows
  \draw [arr]
    (Y) edge (Z) (Z) edge (phi)
    (phi) edge (F.west) (phi) edge (nF.west)
    (F) edge (t1)
    (t1) edge (V0) (t1) edge (V1)
    (V0) edge (Ex)
    (V1) edge (All)
    (Ex) edge (E1.west) (Ex) edge (E2.west) (Ex) edge (E3.west)
    (E1) edge (Ev1) (E2) edge (Ev2) (E3) edge (Ev3)
    (All) edge (A1.west) (All) edge (A2.west) (All) edge (A3.west)
    (A1) edge (Av1) (A2) edge (Av2) (A3) edge (Av3)
    (nF) edge (t2)
    (t2) edge (analog)
    (Y1) edge (Z1) (Z1) edge (C1)
    (Yn) edge (Zn) (Zn) edge (Cn)
    (F) edge (Fv) (nF) edge (nFv)
    (V0) edge (V0v) (V1) edge (V1v)
;
  % dots between ... nodes
  \draw [densely dotted, thick, short=2pt]
    (E1) edge (E2) (E2) edge (E3)
    (A1) edge (A2) (A2) edge (A3)
;
  % back-edges from cloud
  \draw [arr,gray,overlay]
    (analog.east) .. controls ($(Ev1)+(3.7,6.75)$) and ($(Z1)-(0,4.25)$) .. (Z1);   % I bent the arrows so they fit on the page. If you don't agree with this, let us know
  \draw [arr,gray,overlay]
    (analog.east) to [out=-60,in=-90,looseness=.7] (Z);
  \draw [arr,gray,overlay]
    (analog.east) .. controls ($(analog.east)+(2.5,-4.5)$) and ($(Zn)+(0,4)$) .. (Zn);
%
  % back-edges from actual game
  \draw [arr,overlay]
    (E1) edge [out=50,in=-90,looseness=0.7] (Y1)
    (E2) edge [out=40,in=70,looseness=0.8] (Y)
    plot [smooth, tension=0.5] coordinates {(A3.south) ($(analog.east)+(0.1,0.4)$) ($(analog.east)+(-1,-2)$) ($(Yn)+(2,2)$) (Yn.north)}
    %(A3) .. controls ($(analog.east)+(9,-5)$) and ($(Yn)+(0,5)$) .. (Yn)
;
  % omitted back-edges
  \draw [densely dotted, short=2pt]
    (E3) edge ++(-20:1)
    (A1) edge ++(20:1)
    (A2) edge ++(15:1)
;
  % omitted back-edges from C1,Cn
  \draw [densely dotted, short=2pt, overlay]
    (C1.east) to [out=0,in=90,looseness=.5] ($(C1.east)+(1,-1)$)
    (C1.east) to [out=0,in=90,looseness=.5] ($(C1.east)+(.8,-1.1)$)
    (C1.east) to [out=0,in=90,looseness=.5] ($(C1.east)+(.6,-1.2)$)
    (Cn.east) to [out=0,in=-90,looseness=.5] ($(Cn.east)+(1,1)$)
    (Cn.east) to [out=0,in=-90,looseness=.5] ($(Cn.east)+(.8,1.1)$)
    (Cn.east) to [out=0,in=-90,looseness=.5] ($(Cn.east)+(.6,1.2)$)
;
  % initial position
  \draw [arr,overlay] ($(Y)-(1cm,0)$) to (Y);
  \begin{scope}[on background layer]
    \node [draw=none,fill=lightgray!50,rectangle,inner sep=5pt,rounded corners,fit=(Y)(E1)(Av3)(analog)] (rect) {};
    \coordinate (R1) at (Av3.east |- C1);
    \coordinate (Rn) at (Av3.east |- Cn);
    \node [draw=none,fill=lightgray!50,rectangle,inner sep=5pt,rounded corners,fit=(Y1)(C1.north)(C1.south)(R1)] (rect1) {};
    \node [draw=none,fill=lightgray!50,rectangle,inner sep=5pt,rounded corners,fit=(Yn)(Cn.north)(Cn.south)(Rn)] (rectn) {};
  \end{scope}
  % dots on the left
  \draw [loosely dotted, very thick, short=.5cm]
    (Y.east |- rect1.south) edge (Y.east |- rect.north)
    (Y.east |- rectn.north) edge (Y.east |- rect.south)
;
\end{tikzpicture}
\caption{Illustration of the model-checking game $\MC \Gg {v_i}$ for a Büchi game with positions $V = \{v_1,\dots,v_n\}$.
Rounded nodes belong to Verifier, rectangular nodes to Falsifier. Nodes without border are terminal positions representing literals,
dashed nodes are target positions.
}
\label{figModelCheckingFull}
\end{figure*}

To prove \cref{thmSumOfStrategiesTracking}, we show that it follows from a more general sum-of-strategies theorem \cite[Thm.\ 23]{DannertGraNaaTan21}\footnote{A complete proof can be found in the full version \cite[Section~6]{DannertGraNaaTan19}.}.
The general theorem expresses the semiring semantics of arbitrary LFP-formula by means of winning strategies in the corresponding model-checking game;
this does not immediately apply to Büchi games, but it is not difficult to see (and we explain below) that the model-checking game for $\FormulaWin(v)$ on a Büchi game $\Gg$ has the same structure as the original game graph $\Gg$.

Towards the proof, we first sketch the model-checking game $\MC \Gg v$ of $\FormulaWin(v)$ on a Büchi game $\Gg$.
We then briefly recapitulate how semiring values of winning strategies are defined in \cite{DannertGraNaaTan21},
and we show that we can simplify $\MC \Gg v$ without changing these values so that it becomes almost identical to $\Gg$.
This allows us to derive \cref{thmSumOfStrategiesTracking} from the general sum-of-strategies result on $\MC \Gg v$.

\paragraph*{Model-Checking Game}

In general, the model-checking game of an LFP-formula $\psi$ in a structure $\AA$ is a parity game whose positions are pairs of a subformula $\phi(\tx)$ of $\psi$ and an instantiation of the free variables, conveniently written as $\phi(\ta)$ for some tuple $\ta \subseteq \AA$.
The positions belong either to Verifier (who wants to prove $\AA \models \phi(\ta)$) or to Falsifier (who wants to prove $\AA \not\models \phi(\ta)$).
Edges allow the current player to move from a position $\phi(\ta)$ to a direct subformula of $\phi(\ta)$, or from fixed-point literals (such as $Yu$ in our case) back to the entire fixed-point formula.
Terminal positions arise from literals $\phi(\ta)$ and are won by Verifier precisely if $\AA \models \phi(\ta)$.
The model-checking game we are interested in is shown in \cref{figModelCheckingFull}, so we refrain from a complete definition, see e.g.\ \cite[Chap.~4]{AptGraedel11} for more background.

Here we are only concerned with the model-checking game of $\FormulaWin(v)$ in a Büchi game $\Gg$, and we denote this game as $\MC \Gg v$.
Since $\FormulaWin(v)$ only has a single alternation of fixed-point operators, $\MC \Gg v$ has only two priorities and can thus be equivalently represented as a Büchi game (with Verifier as Player~0).
Recall the formula from \cref{subsec:TheFormula} (split into subformulae to ease referencing and with implication already rewritten):
\[
  \FormulaWin(x) \coloneqq{} \big[\gfp Y y.\ [\lfp Z z.\ \phi(Y,Z,z)](y) \big](x),
\]
where
\begin{align*}
  \phi(z) \coloneqq{} &(Fz \land \theta_1(z)) \lor (\neg Fz \land \theta_2(z)), \\[.5\normalbaselineskip]
  \theta_1(z) \coloneqq{} &\big((V_0 z \land \E u (Ezu \land Yu))
  \lor (V_1 z \land \A u (\neg Ezu \lor (Ezu \land Yu)))\big), \\
  \theta_2(z) \coloneqq{} &\big((V_0 z \land \E u (Ezu \land Zu))
  \lor (V_1 z \land \A u (\neg Ezu \lor (Ezu \land  Zu)))\big).
\end{align*}

A depiction of the complete model-checking game $\MC \Gg v$ is shown in \cref{figModelCheckingFull} (with some unavoidable omissions due to space reasons).

\paragraph*{Semiring Values for the Model-Checking Game}

We define strategies for $\MC \Gg v$ as we did for Büchi games, always taking the perspective of Verifier.
To avoid confusion, we use the letter $\Mm$ to denote strategies in the model-checking game, and $\Ss$ for strategies in $\Gg$.
We say that a strategy $\Mm$ is \emph{winning} if on each infinite play, target nodes (i.e., nodes of the form $Y v$) occur infinitely often;
we impose no restrictions on finite plays (which end in literals).
In the following, we only consider strategies for $\MC \Gg v$ that are winning.

Now consider literals, the terminal positions of the model-checking game.
As in \cref{defKInterpretation}, we write $\Lit$ (omitting $V$ and $\tau$) for the set of instantiated literals over the signature of $\FormulaWin$ and the set of positions of $\Gg$.
Given a $\K$-interpretation $\pi$, it assigns to each literal $\lit \in \Lit$ the value $\pi(\lit)$.
Following \cite{DannertGraNaaTan21}, we define the value of a winning strategy $\Mm$ by counting literals.

\begin{defi}
Let $\pi$ be a $\K$-interpretation in an absorptive, fully-continuous semiring.
Let $\Gg$ be a Büchi game and $\Mm$ a winning strategy in $\MC \Gg v$.
For a literal $\lit \in \Lit$, we write $\ecount \Mm \lit \in \N \cup \{\infty\}$ for the number of occurrences
of $\lit$ in $\Mm$ (represented as a strategy tree).
The $\K$-value of $\Mm$ is
\[
    \pi \ext \Mm \coloneqq \prod_{\lit \in \Lit} \pi(\lit)^{\ecount \Mm \lit}.
\]\end{defi}

Notice that for the Boolean interpretation $\pi \colon \Lit \to \Bool$ that maps all literals to truth values according to $\Gg$,
we have $\pi \ext \Mm \neq 0$ if, and only if, $\Mm$ is a winning strategy (in the classical sense) in the standard Boolean model-checking game for $\FormulaWin(v)$ and $\Gg$.
By assigning non-Boolean values for the literals, in particular indeterminates in $\Sinf[X]$, we can track how the literals affect the truth of $\FormulaWin(v)$.
With this terminology, applying the sum-of-strategies theorem in \cite{DannertGraNaaTan21} to $\FormulaWin(v)$ gives:
\begin{thm}
\label{generalSumOfStrategies}
Let $\pi$ be a $\K$-interpretation into an absorptive, fully-continuous semiring.
Then $\pi \ext {\FormulaWin(v)} = \sum \{ \pi \ext \Mm \mid \Mm \text{ is a winning strategy in } \MC \Gg v \}$.
\end{thm}

In order to use this result in our context, we will show that the winning strategies $\Mm$ in $\MC \Gg v$ correspond to winning strategies $\Ss$ in $\Gg$.
For edge-tracking interpretations $\pi$, only the edge literals $Evw$ are relevant and thus the corresponding strategies $\Mm$ and $\Ss$ have the same $\K$-value,
so \cref{generalSumOfStrategies} will then imply \cref{thmSumOfStrategiesTracking}.

\paragraph*{From the Model-Checking Game to the Büchi Game}

Let $\pi$ be an edge-tracking $\K$-interpretation for $\Gg$, so most of the literals are mapped to $0$ or $1$.
We can then remove the corresponding terminal positions in $\MC \Gg v$ and in some cases also their predecessors.
For instance, consider a position $\phi(\ta)$ in $\MC \Gg v$ from which Falsifier can move to a literal $\lit$ with $\pi(\lit) = 0$.
Then every strategy $\Mm$ that visits $\phi(\ta)$ must also visit $\lit$, thus having value $\pi \ext \Mm = 0$,
so we can ignore this strategy for the sum in \cref{generalSumOfStrategies}.
Hence, replacing the position $\phi(\ta)$ by its successor $\lit$ does not change the sum.
On the other hand, if $\pi(\lit) = 1$, then visiting $\lit$ does not affect the $\K$-value of $\Mm$ and hence we can remove $\lit$.
Similar reasoning applies to positions of Verifier.
Moreover, we can always skip over non-target positions with a unique successor, as they neither affect gameplay nor the $\K$-values of strategies.

With these insights, we can simplify the model-checking game in \cref{figModelCheckingFull} quite a bit.
If, say, $v_i \in F$ and $v_i \in V_0$, then the central part of the picture simplifies to:

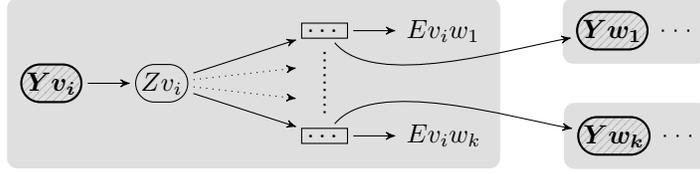
\begin{center}
\begin{tikzpicture}[node distance=1.5cm]
\node [p0,F] (Y) {$Y v_i$};
\node [p0,right=0.7cm of Y] (Z) {$Z v_i$};
\node [p1,minimum height=.2cm,right=of Z,yshift=.7cm] (E1) {\dots};
\node [p1,minimum height=.2cm,right=of Z,yshift=-.7cm] (E2) {\dots};
\node [lit,right=0.7cm of E1] (Ew1) {$Ev_i w_1$};
\node [lit,right=0.7cm of E2] (Ew2) {$Ev_i w_k$};
\draw [dotted, thick, short=5pt] (E1) edge (E2);
\draw [arr] (Y) edge (Z) (Z) edge (E1) (Z) edge (E2);
\draw [arr] (E1) edge (Ew1) (E2) edge (Ew2);
\draw [arr,dotted] (Z) edge ($(E1.west)!0.35!(E2.west)$);
\draw [arr,dotted] (Z) edge ($(E1.west)!0.65!(E2.west)$);
\node [p0,F,right=1.2cm of Ew1] (Y1) {$Y w_1$};
\node [p0,F,right=1.2cm of Ew2] (Y2) {$Y w_k$};
\coordinate [right=.6cm of Y1] (R1);
\coordinate [right=.6cm of Y2] (R2);
\draw [thick,loosely dotted,shorten <=5pt,shorten >=0pt] (Y1) edge (R1) (Y2) edge (R2);
\draw [arr] (E1) edge [out=-50,in=190,looseness=0.7] (Y1);
\draw [arr] (E2) edge [out=50,in=-190,looseness=0.7] (Y2);
\begin{scope}[on background layer]
  \node [draw=none,fill=lightgray!50,rectangle,rounded corners,inner sep=5pt,fit=(Y)(Ew1)(Ew2)] {};
  \node [draw=none,fill=lightgray!50,rectangle,rounded corners,inner sep=5pt,fit=(Y1)(R1)] {};
  \node [draw=none,fill=lightgray!50,rectangle,rounded corners,inner sep=5pt,fit=(Y2)(R2)] {};
\end{scope}
\end{tikzpicture}
\end{center}
Here, $v_i E = \{w_1,\dots,w_k\}$ are the successors of $v_i$ in $\Gg$, so Verifier's moves from $Zv_i$ are only those corresponding to actual edges of $\Gg$.
The other situations are similar, here is the case $v_i \notin F$, $v_i \in V_1$:

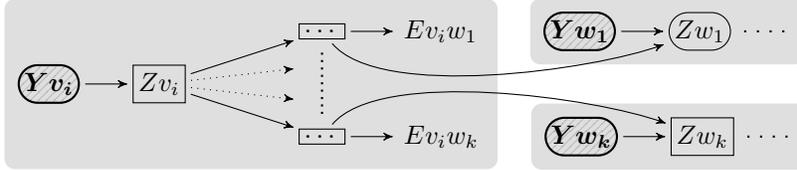
\begin{center}
\begin{tikzpicture}[node distance=1.5cm]
\node [p0,F] (Y) {$Y v_i$};
\node [p1,right=0.7cm of Y] (Z) {$Z v_i$};
\node [p1,minimum height=.2cm,right=of Z,yshift=.7cm] (E1) {\dots};
\node [p1,minimum height=.2cm,right=of Z,yshift=-.7cm] (E2) {\dots};
\node [lit,right=0.7cm of E1] (Ew1) {$Ev_i w_1$};
\node [lit,right=0.7cm of E2] (Ew2) {$Ev_i w_k$};
\draw [dotted, thick, short=5pt] (E1) edge (E2);
\draw [arr] (Y) edge (Z) (Z) edge (E1) (Z) edge (E2);
\draw [arr] (E1) edge (Ew1) (E2) edge (Ew2);
\draw [arr,dotted] (Z) edge ($(E1.west)!0.35!(E2.west)$);
\draw [arr,dotted] (Z) edge ($(E1.west)!0.65!(E2.west)$);
\node [p0,F,right=.8cm of Ew1] (Y1) {$Y w_1$};
\node [p0,F,right=.8cm of Ew2] (Y2) {$Y w_k$};
\node [p0,right=.7cm of Y1] (Z1) {$Z w_1$};
\node [p1,right=.7cm of Y2] (Z2) {$Z w_k$};
\draw [arr] (Y1) edge (Z1) (Y2) edge (Z2);
\coordinate [right=.8cm of Z1] (R1);
\coordinate [right=.8cm of Z2] (R2);
\draw [thick,loosely dotted,shorten <=5pt,shorten >=0pt] (Z1) edge (R1) (Z2) edge (R2);
\draw [arr] (E1) edge [out=-50,in=200,looseness=0.6] (Z1);
\draw [arr] (E2) edge [out=50,in=-200,looseness=0.6] (Z2);
\begin{scope}[on background layer]
  \node [draw=none,fill=lightgray!50,rectangle,rounded corners,inner sep=5pt,fit=(Y)(Ew1)(Ew2)] {};
  \node [draw=none,fill=lightgray!50,rectangle,rounded corners,inner sep=5pt,fit=(Y1)(R1)] {};
  \node [draw=none,fill=lightgray!50,rectangle,rounded corners,inner sep=5pt,fit=(Y2)(R2)] {};
\end{scope}
\end{tikzpicture}
\end{center}
where again $\{w_1,\dots,w_k\}$ are the successors of $v_i$ in $\Gg$.
Notice that $Z v_i$ belongs to Falsifier (as a result of skipping several positions).
In general, $Z v_i$ belongs to Verifier precisely if $v_i$ belongs to Player~0 in $\Gg$.

We can now identify the entire subgraph from $Y v_i$ up to the edge literals $E v_i w_j$ (that is, the gray rectangle) with the position $v_i$ in $\Gg$,
and we call this the \emph{gadget for $v_i$}.
Indeed, if $v_i \in V_0$ then Verifier chooses a successor $w \in v_iE$ and moves to the corresponding gadget,
in analogy to Player~0 choosing a successor in $\Gg$.
Similarly, Falsifier chooses a successor if it is Player~1's turn in $\Gg$.

What remains to discuss are the target positions.
Each gadget has two entry points $Y v$ and $Z v$, and only $Y v$ is a target position.
Notice that when we move from a gadget for $v$ to a gadget for $w$, we use the entry point $Y w$ if, and only if, $v \in F$ (where $F$ is the target set of the original game $\Gg$).
Hence any play that visits infinitely many target positions $Y w$ also visits infinitely many gadgets for positions $v \in F$ (the predecessors of $w$).
We can thus change the target set without affecting the winning strategies:
Instead of the positions $Y w$, we set the target set to $\{ Z v \mid v \in F \}$.
The positions $Y w$ are then regular positions with unique successors and can thus be removed.
As an example, say we have $v_i \in F$, $w_1 \in F$ and $w_k \notin F$. The previous picture then becomes:

\begin{center}
\begin{tikzpicture}[node distance=1.5cm]
\node [p1,F] (Z) {$Z v_i$};
\node [p1,minimum height=.2cm,right=of Z,yshift=.7cm] (E1) {\dots};
\node [p1,minimum height=.2cm,right=of Z,yshift=-.7cm] (E2) {\dots};
\node [lit,right=0.7cm of E1] (Ew1) {$Ev_i w_1$};
\node [lit,right=0.7cm of E2] (Ew2) {$Ev_i w_k$};
\draw [dotted, thick, short=5pt] (E1) edge (E2);
\draw [arr] (Z) edge (E1) (Z) edge (E2);
\draw [arr] (E1) edge (Ew1) (E2) edge (Ew2);
\draw [arr,dotted] (Z) edge ($(E1.west)!0.35!(E2.west)$);
\draw [arr,dotted] (Z) edge ($(E1.west)!0.65!(E2.west)$);
\node [p0,F,right=1.5cm of Ew1] (Z1) {$Z w_1$};
\node [p1,right=1.5cm of Ew2] (Z2) {$Z w_k$};
\coordinate [right=.6cm of Z1] (R1);
\coordinate [right=.6cm of Z2] (R2);
\draw [thick,loosely dotted,shorten <=5pt,shorten >=0pt] (Z1) edge (R1) (Z2) edge (R2);
\draw [arr] (E1) edge [out=-50,in=200,looseness=0.6] (Z1);
\draw [arr] (E2) edge [out=50,in=-200,looseness=0.6] (Z2);
\begin{scope}[on background layer]
  \node [draw=none,fill=lightgray!50,rectangle,rounded corners,inner sep=5pt,fit=(Z)(Ew1)(Ew2)] {};
  \node [draw=none,fill=lightgray!50,rectangle,rounded corners,inner sep=5pt,fit=(Z1)(R1)] {};
  \node [draw=none,fill=lightgray!50,rectangle,rounded corners,inner sep=5pt,fit=(Z2)(R2)] {};
\end{scope}
\end{tikzpicture}
\end{center}

\paragraph*{Proof of \protect{\Cref{thmSumOfStrategiesTracking}}}

Let $\MCP \Gg v$ be the game that results from $\MC \Gg v$ by applying all of the above-mentioned simplifications.
Hence $\MCP \Gg v$ contains for each position $v$ a gadget with unique entry point $Z v$
that belongs to Verifier exactly if $v \in V_0$, and $Z v$ is a target position exactly if $v \in F$. Moreover, the gadget for $v$ is directly
connected to the gadget for $w$ if, and only if, the edge $vw$ exists in $\Gg$.
It is now easy to see that every winning strategy $\Ss$ in the original Büchi game $\Gg$
induces a unique winning strategy $\Mm$ in $\MCP \Gg v$: whenever $\Ss$ visits a position $v$,
then $\Mm$ visits the gadget for $v$ (via the unique entry point $Z v$).
Conversely, every winning strategy $\Mm$ uniquely induces a winning strategy $\Ss$ (which moves to $v$ when $\Mm$ enters the gadget for $v$),
and we thus say that $\Mm$ and $\Ss$ are \emph{corresponding} winning strategies.

\begin{lem}
Let $\Mm$ be a winning strategy in $\MCP \Gg v$.
Let $\Ss$ be a winning strategy in $\Gg$ so that $\Mm$ and $\Ss$ are corresponding strategies.
Then, $\pi \ext \Mm = \pi \ext \Ss$.
\end{lem}
\begin{proof}
As we removed all other literals from $\MC \Gg v$, the only literals occurring in $\Mm$ are edge literals of the form $Evw$,
so $\pi \ext \Mm = \prod_{vw \in E} \pi(Evw)^{\ecount \Mm {Evw}}$.
An edge literal $Evw$ occurs in $\Mm$ whenever $\Mm$ transitions from the gadget for $v$ to the gadget for $w$,
and this happens whenever the edge $vw$ occurs in $\Ss$.
So $\pi \ext \Mm = \prod_{vw \in E} \pi(Evw)^{\ecount \Ss {vw}} = \pi \ext \Ss$.
\end{proof}

This closes the proof of the Sum-of-Strategies Theorem:
since our modifications did not affect the sum over all winning strategies, \cref{generalSumOfStrategies} implies
\begin{align*}
  &\pi \ext {\FormulaWin(v)} \\
  {}={} &\sum \{ \pi \ext \Mm \mid \Mm \text{ is a winning strategy in } \MC \Gg v \} \\
  {}={} &\sum \{ \pi \ext \Mm \mid \Mm \text{ is a winning strategy in } \MCP \Gg v \} \\
  {}={} &\sum \{ \pi \ext \Ss \mid \Ss \text{ corresponds to a winning strategy in } \MCP \Gg v \} \\
  {}={} &\sum \{ \pi \ext \Ss \mid \Ss \in \WinStrat_\Gg(v) \},
\end{align*}
and restricting the sum to winning strategies $\Ss$ that are absorption-dominant from $v$ does not change the overall value. \hfill $\qed$
\endgroup

\section{Case Study: Applications of Semiring Semantics}
\label{sec:Applications}

We now have all of the necessary groundwork to consider applications of semiring semantics for Büchi games.
This section discusses what information the Sum-of-Strategies Theorem provides about winning strategies, how semiring semantics helps to find minimal repairs and why it is not well suited for cost computations.

\subsection{Strategy Analysis}
\label{sec:StrategyTracking}

We begin with the question what information we can derive from the Sum-of-Strategies Theorem.
To this end, we fix a Büchi game $\Gg$ and focus on the $\Sinf[X]$-interpretation $\pitrack$ with $X = \{ X_{uv} \mid u,v \in \Gg \}$.
The values $\pitrack \ext \Ss$ are monomials and we can read off the number of occurrences of each edge in $\Ss$ from the exponents, i.e., the monomial is a representation of the edge profile $\ep \Ss$.
In particular, $\pitrack \ext {\Ss_1} \absorb \pitrack \ext {\Ss_2}$ if,
and only if, $\Ss_1 \absorb \Ss_2$.
The fact that absorptive polynomials are always finite \cite{DannertGraNaaTan21} is thus another way to see that the number of absorption-dominant strategies is finite.

What can we learn from the polynomial $\pitrack \ext {\FormulaWin(v)}$?
First, $\pitrack \ext {\FormulaWin(v)} \neq 0$ holds if, and only if, Player~0 has a winning strategy from $v$.
By \cref{thmSumOfStrategiesTracking}, we can further derive information about all absorption-dominant strategies.
More precisely, we learn which edges each absorption-dominant strategy uses and how often they appear in the strategy tree.
Knowing the edge profile immediately reveals whether the strategy is positional and what the positional choices are.
By counting monomials, we can thus count the positional strategies, as well as the absorption-dominant strategies up to absorption-equivalence.

We can further answer questions such as: can Player~0 still win if we remove edge $e$?
This is the case if, and only if, the polynomial $\pitrack \ext {\FormulaWin(v)}$ contains a monomial without the variable $X_e$ (if there is a winning strategy without $e$, then there is also an absorption-dominant strategy and hence a monomial without $X_e$).
Going further, a more interesting question is: can Player~0 still win if edge $e$ may only be used finitely often in each play?
The answer is not immediately obvious.
Consider for example the strategy $\Ss$ in \cref{fig:RunningStrategy}.
The edge $k$ occurs infinitely often in the strategy tree and we get $\pitrack \ext \Ss = abcefh g^\infty k^\infty m^\infty$.
However, $k$ is clearly played only once in each play consistent with $\Ss$, whereas edge $m$ is played infinitely often.
We cannot distinguish edges $k$ and $m$ just from $\pitrack \ext \Ss$, but we can do so if we compute $\pitrack \ext {\FormulaWin(w)}$ for all positions $w \in V$, by the following criterion (notice that all of these values are computed anyway for the fixed-point iteration).

\begin{prop}
Let $\Ss \in \WinStrat_\Gg(v)$ be absorption-domi\-nant from $v$, and let $e = uw \in E$ be an edge with $\ecount \Ss e = \infty$.
Then there is a unique (positional) strategy $\Ss_w \in \WinStrat_\Gg(w)$ such that $\pitrack \ext {\Ss_w} \absorb \pitrack \ext \Ss$.
Moreover, $\Ss$ admits a play in which $e$ occurs infinitely often if, and only if, $e$ occurs
in $\Ss_w$.
\end{prop}

\begin{proof}
Consider the strategy tree $\Ss$ and let $\rho w$ be an occurrence of $w$ in $\Ss$.
By assumption, $w$ occurs infinitely often in $\Ss$.
But then, for all successors $\rho w w'$ of $\rho w$, also $w'$ occurs infinitely often in $\Ss$.
Indeed, either $w \in V_1$ and every occurrence of $w$ must be followed by an occurrence of $w'$;
or $w \in V_0$ and $\Ss$ plays positionally from $w$ by \cref{stratInfinitePositional}, so again every occurrence of $w$ is followed by an occurrence of $w'$ in $\Ss$.
By induction, it follows that $\Ss$ plays positionally from $w$ and from all positions occurring below $\rho w$.
In particular, the substrategy $\Ss_w$ that $\Ss$ plays from $\rho w$ (and any other occurrence of $w$) is positional.
As a substrategy, we trivially have $\pitrack \ext {\Ss_w} \absorb \pitrack \ext \Ss$.
To see that $\Ss_w$ is unique with this property, notice that a strategy $\Ss' \in \Strat_\Gg(w)$ deviating from $\Ss_w$ must play an edge which does not occur in $\Ss_w$ and hence also not in $\Ss$, so $\pitrack \ext {\Ss'} \not\absorb \pitrack \ext \Ss$.

For the second statement, first assume that $\Ss$ admits a play $v v_1 v_2 v_3 \dots$ in which the edge $e=uw$ occurs infinitely often.
Let $v_i = w$ be the first occurrence of $w$.
Then the remaining play $v_i v_{i+1} v_{i+2} \dots$ is consistent with $\Ss_w$ and hence $e$ occurs (infinitely often) in $\Ss_w$.
Conversely, assume that $e$ occurs in $\Ss_w$, so there is some $\rho \in V^*$ such that $w \rho u w \in \Ss_w$.
As $\Ss_w$ is positional, we can repeat $\rho u w$ to obtain the infinite play $w (\rho u w)^\omega$ consistent with $\Ss_w$.
And since $\Ss_w$ is a substrategy of $\Ss$, this induces an infinite play of the form $\rho' w (\rho u w)^\omega$ in $\Ss$ which indeed uses the edge $uw$ infinitely often.
\end{proof}

\begin{example}
Consider the strategy $\Ss$ in \cref{fig:RunningStrategy} with $\pitrack \ext \Ss = abcefh g^\infty k^\infty m^\infty$ and the edge $k$ from $u$ to $w$.
Since edge $n$ does not occur in $\pitrack \ext \Ss$, the only winning strategy $\Ss_w$ from $w$ we need to consider is the strategy that always stays at $w$, with $\pitrack \ext {\Ss_w} = m^\infty \absorb \pitrack \ext \Ss$.
As $k$ does not occur in $\Ss_w$, we conclude that it occurs only finitely often (and hence at most once) in each play consistent with $\Ss$.

If, on the other hand, we consider edge $m$ (which also leads to position $w$), we see that $m$ occurs in $S_w$ and we can thus infer that $\Ss$ contains a play visiting $m$ infinitely often.
\end{example}

Summarizing the results of this section, we see that semiring semantics in $\Sinf[X]$ is very informative and allows us to derive important information about the winning strategies.

\begin{cor}\label{corPolynomialInformation}
From the polynomial $\pitrack \ext {\FormulaWin(v)}$, we can efficiently (in the size of the polynomial) derive the following information:
\begin{itemize}
\item whether Player $0$ wins from $v$,
\item the edge profiles of all absorption-dominant winning strategies from $v$,
\item the number and precise shape of all positional winning strategies from $v$,
\item whether Player $0$ can still win from $v$ if only a subset of the edges is allowed.
\end{itemize}

Given the polynomials $\pitrack \ext {\FormulaWin(v)}$, for \emph{all} positions $v$,
we can further derive for each (equivalence class of an) absorption-dominant strategy and each edge, how often the edge can occur in a play consistent with the strategy.
\end{cor}

\subsection{Reverse Analysis}
\label{sec:ReverseAnalysis}

Instead of tracking strategies in a fixed game, we may also ask questions such as:
assuming Player~1 wins from $v$, what are minimal modification to $\Gg$ such that instead Player~0 wins?
The generality of semiring semantics enables us to answer such questions by choosing appropriate semirings and interpretations.

More precisely, let $\Gg = (V,V_0,V_1,E,F)$ be a Büchi game and $v \in \Gg$ a position from which Player~1 wins.
Let $E^- \subseteq E$ and $E^+ \subseteq V^2 \setminus E$ be sets of edges we are allowed to delete or add, respectively.
We call a set of edges in $E^\pm \coloneqq E^- \cup E^+$ a \emph{repair} if Player~0 wins when these edges are deleted or added.
Our goal is to determine all (preferably minimal) repairs.
We achieve this by evaluating $\FormulaWin(v)$ in a modified polynomial semiring, similar to the computation of repairs for database queries in \cite{XuZhaAlaTan18},
except that here we need absorptive polynomials to deal with fixed points.

\bigskip\noindent\textbf{Dual-Indeterminates.}
To track negative information, such as the absence of an edge, we follow the approach in \cite{GraedelTan17,XuZhaAlaTan18,DannertGraNaaTan21} and extend our semiring by dual-indeterminates $\nnX = \{ \nnx \mid x \in X \}$.
The idea is to label a literal and its negation by corresponding indeterminates $x$ and $\nnx$.
We must then avoid monomials such as $x\nnx$, as they represent contradictory information.
To this end, we consider the quotient of $\Sinf[X \cup \nnX]$ with respect to the congruence generated by $x \cdot \nnx = 0$ for $x \in X$
and refer to the resulting quotient semiring as \emph{dual-indeterminate absorptive polynomials} $\Sinf[X,\nnX]$.
This semiring inherits most of the properties of $\Sinf[X]$.
Most importantly, any assignment $h \co X \cup \nnX \to \K$ into an absorptive, fully-continuous semiring $\K$ that respects dual-indeterminates, i.e., $h(x) \cdot h(\nnx) = 0$, lifts to a fully-continuous homomorphism analogous to \cref{universality}.

We then replace $\pitrack$ by an $\Sinf[X,\nnX]$-interpretation $\pirev^\pm$ with $X = \{ X_e \mid e \in E^\pm\}$:
if $vw \in E^\pm$, we set $\pirev^\pm(Evw) = X_{vw}$ and $\pirev^\pm(\neg Evw) = \smash{\nn{X_{vw}}}$,
all other literals are mapped to $0$ or $1$ according to $\Gg$ (cf.\ \cref{fig:Interpretations}).
Notice that $\pirev^\pm$ is neither model-defining nor edge-tracking, but still satisfies $\pirev^\pm(\lit) \bcdot \pirev^\pm(\neg \lit) = 0$ for all literals $\lit$.

\bigskip\noindent\textbf{Back and Forth between Monomials and Models.}
Let $X^\pm = \{ X_e \mid e \in E^+ \} \cup \{ \nn{X_e} \mid e \in E^- \}$.
Given $Y \subseteq X^\pm$, we further write $E(Y) = \{ e \mid X_e \in Y \text{ or } \nn{X_e} \in Y \}$ for the set of edges mentioned in $Y$.
We denote the set of all (dual-)indeterminates occurring in a monomial $m$ by $\var(m) = \{ x \in X \cup \nnX \mid m(x) > 0 \}$.
By examining what combinations of indeterminates from $X^\pm$ occur in the monomials of $\pirev^\pm \ext {\FormulaWin(v)}$, we can read off all minimal repairs as follows.

\begin{prop}\label{reverseBothDirections}
In the above setting, the following holds:
\begin{enumerate}
\item 
Let $m \in \pirev^\pm \ext {\FormulaWin(v)}$ be a monomial.
Then the set $E(\var(m) \cap X^\pm)$ is a repair.

\item
Let $R \subseteq E^\pm$ be a repair.
Then there is a monomial $m \in \pirev^\pm \ext {\FormulaWin(v)}$ such that
$E(\var(m) \cap X^\pm) \subseteq R$.
If $R$ is minimal, then $E(\var(m) \cap X^\pm) = R$.
\end{enumerate}
\end{prop}

Before proving \cref{reverseBothDirections}, we illustrate the computation of minimal repairs in a small example.

\begin{example}
In the following game, Player~1 wins from $v$.
We are interested in the minimal repairs with $E^+ = \{c\}$ and $E^- = \{a,b\}$.
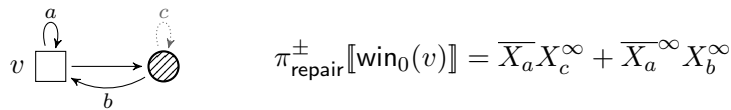
\begin{center}
\begin{tikzpicture}[node distance=1.5cm, baseline]
\node [p1,label={left:$v$}] (0) {};
\node [p0,F,right of=0] (1) {};
\draw [arr]
    (0) edge (1)
    (1) edge [bend left] node {$b$} (0)
    (1) edge [loop above,densely dotted,gray!80!black] node {$c$} (1)
    (0) edge [loop above] node {$a$} (0)
    ;
\end{tikzpicture}
\hspace{1cm}
$\displaystyle \pirev^\pm \ext {\FormulaWin(v)} = \nn{X_a} X_c^\infty + \nn{X_a}^\infty X_b^\infty$\\[.3cm]
\end{center}
Evaluating $\FormulaWin(v)$ in the $\Sinf[X,\nnX]$-interpretation $\pirev^\pm$ described above results in two monomials.
The first yields the repair $\{a,c\}$, the second yields the minimal repair $\{a\}$ (notice that $X_b \notin X^\pm$, as edge $b$ is already present).
The reason why we get two monomials is that we track also positive usage of edge $b$ by $X_b$, but are only interested in the negative indeterminate $\nn{X_b}$ for the repairs.
\end{example}

\begin{proof}[Proof of \cref{reverseBothDirections}]
We prove both statements by considering homomorphisms into the Boolean semiring $\Bool$.
For the first statement, let $m \in \pirev^\pm \ext {\FormulaWin(v)}$ be a monomial and
let $h \colon X \cup \nnX \to \Bool$ be the unique function that respects dual-indeterminates and satisfies
\begin{itemize}
\item $h(x) = \Bone$, for all $x \in \var(m)$
\item $h(X_e) = \Bzero$, if $X_e, \nn{X_e} \notin \var(m)$ and $e \in E^+$ (do not add $e$ without reason),
\item $h(X_e) = \Bone$, if $X_e, \nn{X_e} \notin \var(m)$ and $e \in E^-$ (do not remove $e$ without reason).
\end{itemize}

Then, $h$ lifts to a fully-continuous semiring homomorphism $h \colon \Sinf[X,\nnX] \to \Bool$ with $h(m) = \Bone$.
Moreover, $h \circ \pirev^\pm$ is a Boolean interpretation which corresponds to a Boolean model $\Gg'$.
Since semiring semantics are preserved by fully-continuous homomorphisms, we have $h \circ \pirev^\pm \ext {\FormulaWin(v)} = h(\pirev^\pm \ext {\FormulaWin(v)}) \ge h(m) = \Bone$ and hence $\Gg' \models \FormulaWin(v)$.
By the choice of $h$, the model $\Gg'$ is equal to $\Gg$ except that we add all edges $e \in X^+$ with $X_e \in \var(m)$, and remove all $e \in X^-$ with $\nn{X_e} \in \var(m)$.
Hence $\Gg'$ results from $\Gg$ by adding or deleting the edges $E(\var(m) \cap X^\pm)$, and since $\Gg' \models \FormulaWin(v)$, this set is a repair as claimed.

For the second statement, let $R \subseteq E^\pm$ be a repair and consider the repaired game $\Gg' \models \FormulaWin(v)$.
As $\Gg'$ differs from $\Gg$ only by edges in $E^\pm$, there is a unique assignment $h \colon X \cup \nnX \to \Bool$ such that $h \circ \pirev^\pm$ corresponds to $\Gg'$.
Again, $h$ lifts to a fully-continuous homomorphisms and we thus get $\Bone = h \circ \pirev^\pm \ext {\FormulaWin(v)} = h( \pirev^\pm \ext {\FormulaWin(v)})$.
So there must be a monomial $m \in \pirev^\pm \ext {\FormulaWin(v)}$ with $h(m) = \Bone$.
Consider the set $\var(m) \cap X^\pm$.
If $X_e \in \var(m) \cap X^\pm$, then $h(X_e) = \Bone$ and hence $e \in R$ by construction of $h$.
Further, $\nn{X_e} \in \var(m) \cap X^\pm$ implies $h(\nn{X_e}) = \Bone$ and thus again $e \in R$ by construction of $h$.
This proves $E(\var(m) \cap X^\pm) \subseteq R$.
If $R$ is minimal, we have equality:
otherwise $E(\var(m) \cap X^\pm)$ would be a smaller repair by the first statement, contradicting minimality.
\end{proof}

We remark that these results ignore the exponents of the monomials, so we can drop exponents from $\Sinf[X,\nnX]$ (in other words, we use exponents from $\Bool$ instead of $\N \cup \{\infty\}$) and work in the resulting, simpler semiring $\PosBool[X,\nnX]$ (the dual-indeterminate quotient of the semiring $\PosBool[X]$, see e.g.\ \cite{GreenTan17}).

\subsection{Target synthesis}

The reverse analysis approach and the proof technique based on homomorphisms are not limited to questions about edges, but are general concepts of semiring provenance analysis.
As an example of a different application, we consider the synthesis of the target set $F$.
More precisely, we consider a game arena with positions $V = V_0 \dcup V_1$ and edges $E$ and want to compute all minimal choices for the set $F$ so that Player~0 wins the resulting Büchi game from some fixed starting position $u \in V$.

Similar to the computation minimal repairs, we can solve this task with an interpretation $\piwin$ over the dual-indeterminate semiring $\PosBool[X,\nnX]$ which interprets most literals, including edge literals, by Boolean values, but tracks the target set $F$ using corresponding pairs of dual-indeterminates $\piwin(Fv) = X_v$ and $\piwin(\neg Fv) = \nn{X_v}$ for each position $v \in V$ (cf. \cref{fig:Interpretations}).
We can then derive all possible minimal choices for $F$ from the polynomial $\piwin \ext {\FormulaWin(u)}$, as illustrated in the following example.

\begin{example}
Consider the following game arena.
What are the minimal choices of the target set $F$ so that Player~0 wins from position $a$, or $b$?\\
\begin{center}
\begin{tikzpicture}[node distance=1.5cm, baseline]
\node [p0,label={below:$\vphantom{b}a$}] (0) {};
\node [p1,right of=0,label={below:$b$}] (1) {};
\node [p1,right of=1,label={below:$\vphantom{b}c$}] (2) {};
\draw [arr]
    (0) edge [loop above] (0)
    (2) edge [loop above] (2)
    (0) edge [bend left=10pt] (1)
    (1) edge [bend left=10pt] (0)
    (1) edge (2)
    (2) edge [bend right] (0);
\end{tikzpicture}
\hspace{1cm}
$\displaystyle \begin{aligned}
\piwin \ext {\FormulaWin(a)} &= X_a + \nn{X_a} X_b X_c \\
\piwin \ext {\FormulaWin(b)} &= X_a X_b X_c + X_a \nn{X_b} X_c + \nn{X_a} X_b X_c 
\end{aligned}$\\[.3cm]
\end{center}
Using the $\PosBool[X,\nnX]$-interpretation $\piwin$, we can derive that to win from $a$, the target set must contain at least $\{a\}$ or $\{b,c\}$.
To win from $b$, it must contain at least $\{a,b,c\}$, $\{a,c\}$ or $\{b,c\}$ (notice that $\{a,b,c\}$ is a valid choice for $F$, but not minimal due to the presence of negative indeterminates).
This covers all minimal possible choices of the target set.
\end{example}

In general, the positive indeterminates $X_v$ in each monomial of $\piwin \ext {\FormulaWin(u)}$ induce one possible choice of the target set $F$, and conversely every minimal choice of $F$ occurs as a monomial; this follows by the same arguments as in \cref{reverseBothDirections}.

We remark that we can do slightly better, as we do not need to track negative information for the synthesis problem and hence do not need the negative indeterminates $\nn{X_v}$.
In the above example, we would then obtain the polynomials $X_a + X_b X_c$ and $X_a X_c + X_b X_c$ for positions $a$ and $b$, respectively, which correspond exactly to the minimal choices for $F$.
This can be achieved by setting $\piwin(\neg Fv) = 1$ for all $v \in V$ and observing that this has the same effect as omitting the subformula $\neg F v$ from $\FormulaWin$.
Since the resulting formula is equivalent in Boolean semantics, the reasoning in the proof of \cref{reverseBothDirections} can be adapted to this setting.
However, here we presented the general dual-indeterminate approach which does not depend on the actual formula we consider and hence also works in many other scenarios of provenance analysis, beyond the analysis of Büchi games.

\subsection{Complexity}

The previous applications show that once we have computed the polynomial $\pitrack \ext {\FormulaWin(v)}$ or $\pirev^\pm \ext {\FormulaWin(v)}$,
it is easy to derive information about strategies or minimal repairs -- but how efficient is the computation of the polynomial in the first place?
Recall that for a fixed formula, the LFP model-checking problem for a classical structure can be solved in polynomial time in the size of the structure (here the number of positions).
The analogous problem in semiring semantics, to compute the value $\pitrack \ext {\FormulaWin(v)}$, is more involved since depending on the semiring, fixed-point iterations can be infinite, and contrary to the Boolean case, it is not
trivial that fixed points can be computed efficiently.
However, it has been shown in \cite{Naaf21} that in absorptive, fully-continuous semirings, a least or greatest fixed point of a polynomial system, such as the ones induced by LFP-formulae, can be computed using only a polynomial number of semiring operations (including the infinitary power operation).
By evaluating nested fixed points recursively, we can thus compute the semiring value of a fixed\footnote{In general, this applies to all LFP formulae in which the alternation depth, the arity of fixed-point relations and the number of free variables in any subformula is bounded by a constant.} formula in a number of semiring operations that is polynomial in the size of the structure (here in the number of positions).

Whether this computation is efficient depends on the complexity of the semiring operations and hence on the semiring.
Indeed, although the \emph{number} of semiring operations is polynomial, the resulting polynomial $\pitrack \ext {\FormulaWin(v)}$ can nevertheless have an exponential number of monomials.
This is, in general, unavoidable as both the number of (positional) winning strategies 
as well as the number of minimal repairs can be exponential in the size of the game.
For instance, games of the following form have an exponential number of positional (and hence absorption-dominant) winning strategies:
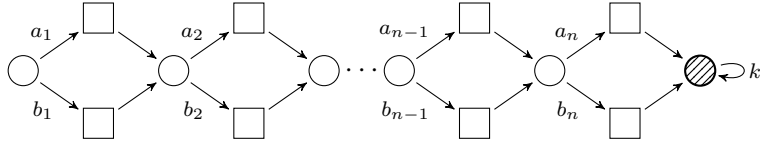
\begin{center}
\begin{tikzpicture}[node distance=1cm]
\node [p0] (0) {};
\node [p1,right of=0,yshift=7mm] (1a) {};
\node [p1,right of=0,yshift=-7mm] (1b) {};
\node [p0,right of=1a,yshift=-7mm] (2) {};
\node [p1,right of=2,yshift=7mm] (3a) {};
\node [p1,right of=2,yshift=-7mm] (3b) {};
\node [p0,right of=3a,yshift=-7mm] (4) {};
\node [p0,right of=4] (4') {};
\node at ($(4)!0.5!(4')$) {$\,\cdots$};
\node [p1,right of=4',yshift=7mm] (5a) {};
\node [p1,right of=4',yshift=-7mm] (5b) {};
\node [p0,right of=5a,yshift=-7mm] (6) {};
\node [p1,right of=6,yshift=7mm] (7a) {};
\node [p1,right of=6,yshift=-7mm] (7b) {};
\node [p0,F,right of=7a,yshift=-7mm] (8) {};
\draw [arr]
    (0) edge node {$a_1$} (1a) (0) edge node [swap] {$b_1$} (1b)
    (1a) edge (2) (1b) edge (2)
    (2) edge node {$a_2$} (3a) (2) edge node [swap] {$b_2$} (3b)
    (3a) edge (4) (3b) edge (4)
    (4') edge node {$a_{n-1}$} (5a) (4') edge node [swap] {$b_{n-1}$} (5b)
    (5a) edge (6) (5b) edge (6)
    (6) edge node {$a_n$} (7a) (6) edge node [swap] {$b_n$} (7b)
    (7a) edge (8) (7b) edge (8)
    (8) edge [loop right] node {$k$} (8)
;
\end{tikzpicture}
\end{center}

Algorithmically, the computation of the polynomials $\pitrack \ext {\FormulaWin(v)}$ 
or $\pirev^\pm \ext {\FormulaWin(v)}$ is thus
infeasible in many cases. It should be noted that the questions addressed in 
\cref{corPolynomialInformation} can be solved by direct methods, and 
just for finding \emph{some} winning strategy or repair, these will  in many cases be more efficient
than computing the polynomials.
Thus, the main benefit of semiring semantics in $\Sinf[X]$ does not lie in a more efficient 
algorithmic method to compute 
some specific winning strategy or repair, but rather in providing a general and compact description of many 
important strategies at once, from which we can directly derive the answers to a number of  different questions 
concerning the strategy analysis of a Büchi game. This is attractive in cases where the relevant
polynomials are reasonably small.

One scenario where we can achieve this is, when only some of the edges are tracked:
instead of the interpretation $\pitrack$ that assigns indeterminates to all edges, we 
map edges we are not interested in to $1$.
The resulting $\Sinf[X]$-interpretation remains edge-tracking, so the Sum-of-Strategies Theorem still applies
and we can see which sets of the tracked edges are at least required for a winning strategy.
In particular, $\FormulaWin(v)$ evaluates to $1$ if there is a winning strategy that avoids all tracked edges.
If this information is sufficient, for instance if we know that a certain part of the game must be visited and only care about edges within this part, we can make provenance analysis more efficient by only tracking small parts of a potentially large game.

\subsection{Limitations}

We have seen that semiring semantics in polynomial semirings such as $\Sinf[X]$ reveals useful information about  strategies.
However, there are also limitations to this framework when it comes to certain typical applications of provenance analysis
such as cost computation. Given cost annotations of the edges, say in the tropical semiring $\Trop=(\mathbb{R}_{+}^{\infty},\min,+,\infty,0)$ (which is absorptive and fully continuous), do semiring semantics provide a sensible cost measure for the evaluation of $\FormulaWin(v)$? Or, more intuitively, for the cost that Player~0 has to pay to win?
For first-order logic and acyclic games, this is indeed the case \cite{GraedelTan17,GraedelTan20}, but here fixed-point computations and infinite plays complicate the situation.
To see this, let $\pi$ be an edge-tracking $\Trop$-interpretation that assigns to each edge $vw$ in $\Gg$ a cost $\pi(Evw) \neq \infty$ (cost $0$ is allowed).
By the Sum-of-Strategies Theorem, we can view $\pi \ext {\FormulaWin(v)}$ as the minimum over the cost of each strategy, and in the tropical semiring, this cost can be expressed as follows (computed over real numbers):
\[
    \pi \ext \Ss = \sum_{vw \in E} {\ecount \Ss {vw}} \cdot \pi(Evw).
\]
That is, Player~0 has to pay for each \emph{occurrence} of an edge in the strategy tree.
While this is certainly a possible cost measure for a strategy, it is debatable whether it is an intuitive one.
Many edges will occur infinitely often in a strategy, and then $\infty \cdot \pi(Evw)$ is either $0$ or $\infty$, and the latter leads to the overall value $\pi \ext \Ss = \infty$.
Even if an edge occurs only once per play, but infinitely often in $\Ss$, Player~0 has to pay the cost $\infty \cdot \pi(Evw)$.
Instead, it might be more intuitive to define the cost of a strategy so that

\begin{enumerate}
\item Player~0 only has to pay once for an edge, no matter how often it occurs (think of an ``unlocking fee''),
\item Player~0 only has to pay for the maximal cost of any play consistent with $\Ss$, but not for all plays simultaneously.
\end{enumerate}

We claim that neither of these options is possible without adapting our notion of semiring semantics.
Notice that we can actually solve (1) by first computing the polynomial $\pitrack \ext {\FormulaWin}$ in $\Sinf[X]$, then dropping exponents (or directly working in $\PosBool[X]$), and then instantiating each variable $X_{vw}$ by its cost to obtain a value in $\Trop$.
But this is only an indirect solution and computing $\pitrack \ext {\FormulaWin}$ may incur an exponential blowup even though we only compute a single cost value.

A direct computation, in the sense that $\pi \ext \Ss$ yields the desired cost value for some suitable $\pi$, is not possible.
To see this for (1), note that we multiply the cost for each occurrence of an edge (in $\Trop$, semiring multiplication is defined as summation on real numbers, but we stick to the general vocabulary).
Say we have two edges $vw$ and $v'w'$ with the same cost $\pi(Evw) = \pi(E v'w') = c$.
To pay only once for $vw$, we would need $\pi(Evw) \bcdot \pi(Evw) = c$, but at the same time $\pi(Evw) \bcdot \pi(Ev'w') = c^2$ for different edges, a contradiction.
The issue is that the information on whether two edges are equal is abstracted away by $\pi$.

A different argument explains why (2) is not possible.
At a position in $V_0$, we want to \emph{minimize} the cost over all possible choices (corresponding to the existential quantification in $\FormulaWin$).
For consecutive edges, we have to \emph{add} costs up, requiring a second operation.
But for positions in $V_1$, to fulfil (2) we must \emph{maximize} the cost over all possible choices, thus requiring a third operation.
Hence this cost measure is not expressible in semirings with only two operations and we would need different algebraic structures.

\section{Conclusion}

Based on a recent line of research on semiring provenance analysis that lead from database theory to semiring semantics for LFP,
we reported here on a case study that puts semiring semantics to use for a strategy analysis in Büchi games.
The choice of Büchi games has been motivated on one side by their relevance for applications in 
the synthesis and verification of reactive systems, on the other side because they provide one of the simplest non-trivial
cases of infinite games for which the definability of winning positions requires an alternation between least and greatest fixed points --
and can thus not be treated by simpler classes of semirings such as the $\omega$-continuous ones 
used for Datalog and reachability games. 

The aim of the case study was to illustrate how semiring semantics can be applied to more complex games,
featuring infinite plays and complicated winning conditions, and what kind of insights it provides (or fails to provide)
about the winning strategies in the game.
This is captured in the central Sum-of-Strategies Theorem and its applications.
This non-trivial result can be seen as a simpler version of the general sum-of-strategies characterization in terms of model-checking games in \cite{DannertGraNaaTan21}
and it essentially identifies the value of the statement that Player~0 wins with the sum of the
valuations of all (absorption-dominant) winning strategies.
While this applies to the class of all absorptive, fully-continuous semirings,
the most important semirings for our analysis are generalized absorptive polynomials $\Sinf[X]$.
Due to their universal property, these provide the most general information,
allowing us to read off the edge profiles of all absorption-dominant strategies.

With this information, we can count positional strategies,
we can determine whether a particular move is needed (once or even infinitely often) for winning the game,
and we can compute minimal ``repairs'' for a game.
The method of semiring valuations is rather flexible; we can use different semirings than $\Sinf[X]$, and we can tailor the set of moves that 
we track to make the resulting polynomial smaller and its computation more efficient.
Of course, not all relevant questions about strategies can be answered directly by semiring valuations,
and as an example for a limitation of this method, we have shown that minimal cost computations
provide serious obstacles to a semiring treatment.

The Sum-of-Strategies Theorem motivates the notion of absorption-dominant strategies which is of interest in its own right as it
captures strategies that are minimal with respect to the multiplicities of the edges they use.
To understand these strategies, we discussed how they relate to other classes of simple strategies, namely to positional and persistent strategies,
and we have shown that these form a strict hierarchy.

Finally, we remark that although Büchi games have been chosen as the topic of our case study,
the method of semiring valuations in absorptive semirings is not confined to this case. In principle,
it can be applied to different formulae and generalizes in particular to other games such as parity games,
as long as the winning positions are definable in fixed-point logic.
The win-formula for parity games is more complicated, and is parametrised by the number of priorities,
and so the Sum-of-Strategies Theorem requires different technical 
details, but can be established along the same lines.

%%%%%%%%%%%%%%%%%%%%%%%%%%%%%%%%
%%  REFERENCES
%%%%%%%%%%%%%%%%%%%%%%%%%%%%%%%%

\bibliographystyle{alphaurl}
\bibliography{paper}

\end{document}